\documentclass[prb, twocolumn, superscriptaddress, showpacs,floatfix]{revtex4}  
\usepackage{hyperref}
\usepackage{amssymb}
\usepackage{amsmath}
\usepackage{float}
\usepackage{bbm}
\usepackage{graphicx}
\usepackage{epsfig}
\usepackage{epstopdf}
\usepackage[usenames]{color}

\def\Im{\hbox{Im}}

\begin{document}

\author{Emanuel Gull}
\affiliation{Department of Physics, Columbia University, 538 West 120th Street, New York, New York 10027, USA}
\author{Olivier Parcollet}
\affiliation{Institut de Physique Th{\'e}orique, CEA, IPhT, CNRS, URA 2306, F-91191 Gif-sur-Yvette, France}
\author{Philipp Werner}
\affiliation{Theoretische Physik, ETH Zurich, 8093 Z{\"u}rich, Switzerland}
\author{Andrew J. Millis}
\affiliation{Department of Physics, Columbia University, 538 West 120th Street, New York, NY 10027, USA}

%\title{On the momentum-sector-selective transition in the 8 site dynamical mean-field approximation to the two-dimensional Hubbard model}
\title{Momentum-sector-selective metal-insulator transition in the eight-site
dynamical mean-field approximation to the Hubbard model in two
dimensions}

\date{\today}

\hyphenation{}

\begin{abstract}
We explore the momentum-sector-selective metal insulator  transitions recently found in the eight - site dynamical cluster approximation to the two-dimensional Hubbard model. The phase diagram in the space of interaction and second-neighbor hopping  is established. The initial transitions from Fermi-liquid like to sector-selective phases are found to be of second order,  caused by the continuous  opening of an energy gap whereas the other transitions are found to be of first order.  In the sector-selective phase the Fermi surface regions which are not gapped are found to have a non-Fermi-liquid self-energy. We demonstrate that the phenomenon is not caused by the Van Hove divergence in the density of states. The sector-selective and insulating phases are characterized by a 
cluster spin correlation function that is strongly peaked at the commensurate antiferromagnetic wave vector $(\pi,\pi)$
but the  model has no nematic instability. Comparison to dynamical mean-field studies on smaller clusters is made.
\end{abstract}


\maketitle

\section{Introduction}

The ``pseudogap'', a suppression of the electronic spectral function occurring for momentum states along the Brillouin-zone face but not for states along the zone diagonal, is a basic and still ill-understood feature of hole-doped high-temperature cuprate superconducting materials. (In electron-doped cuprates a different effect, confusingly also sometimes termed as pseudogap, is attributed to the presence of or proximity to long-ranged two-sublattice antiferromagnetic order.)  The pseudogap, which occurs in the absence of any obvious long-ranged order, is a dramatic example of the more general phenomenon of ``momentum-space differentiation,'' an increase in the variation in physical quantities around the Fermi surface as the insulating phase is approached. Its  origin and consequences  remain hotly debated topics. 

\begin{figure}[t]
\includegraphics[angle=0, width=0.43\columnwidth]{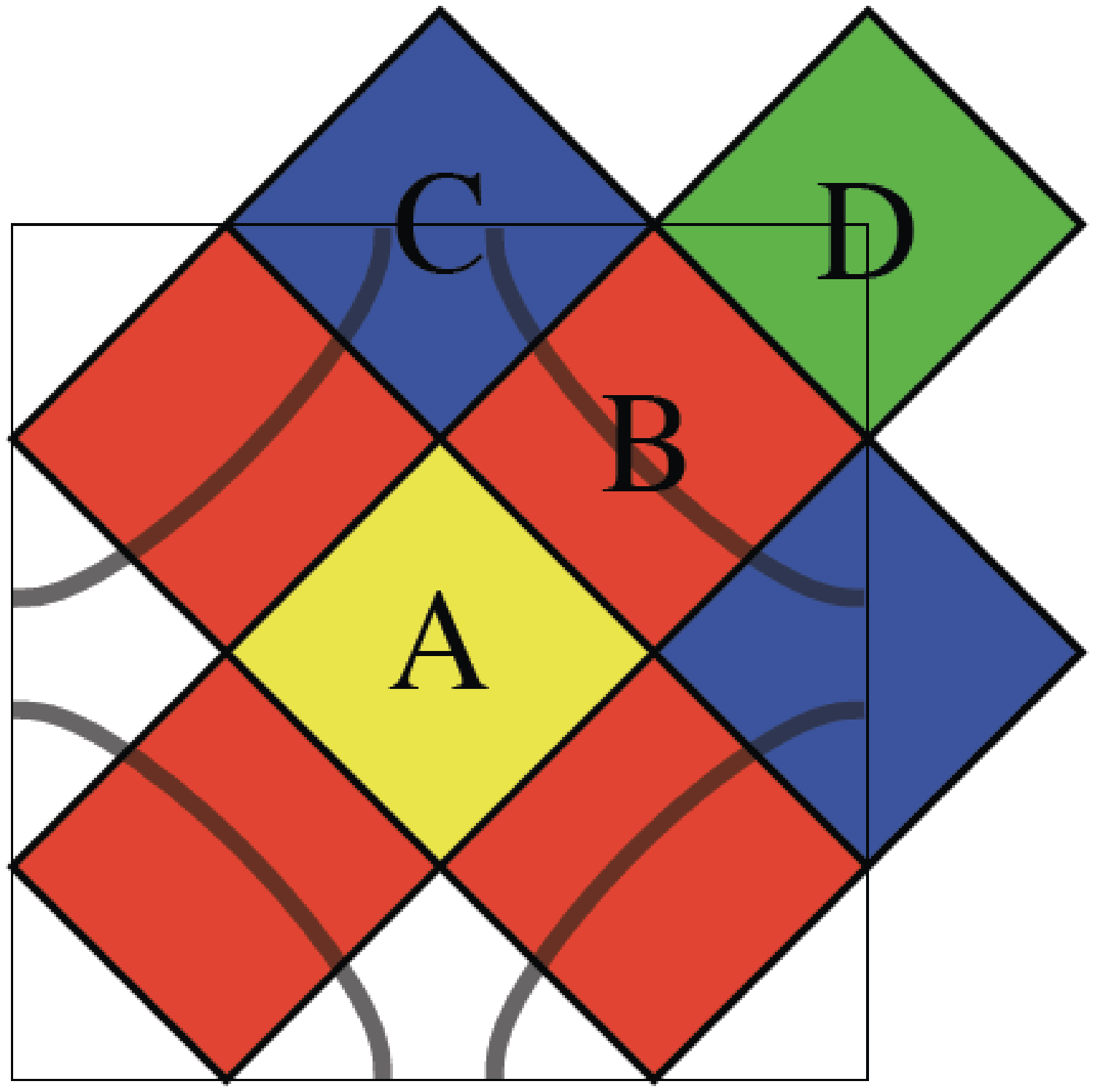}
%\hfill
\includegraphics[angle=0, width=0.52\columnwidth]{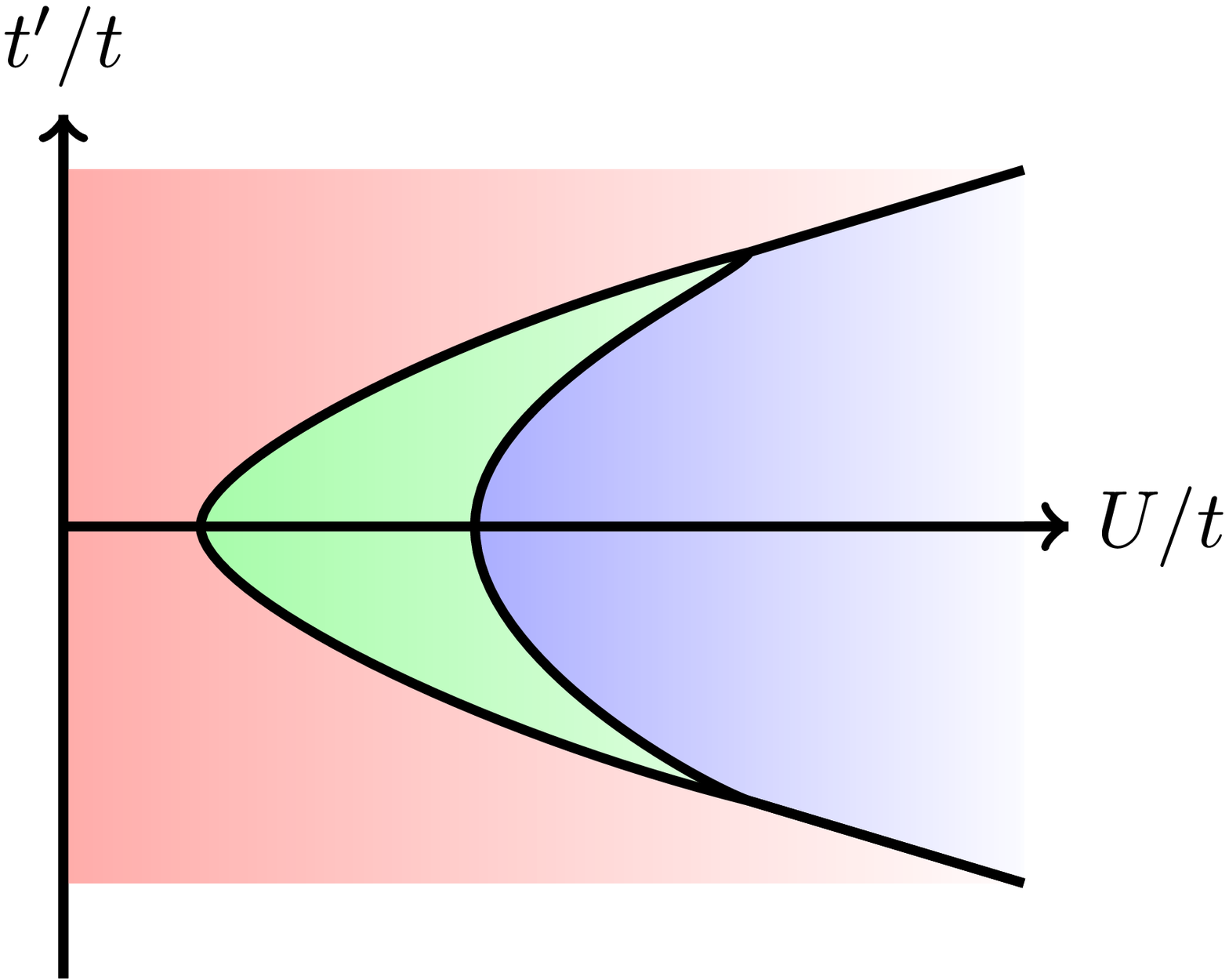}
\caption{
Left panel: 
Brillouin-zone partitioning associated with the eight-site cluster including definition of the four inequivalent momentum sectors $A$, $B$, $C$, and $D$. The non-interacting Fermi surface for $t'=-0.15t$ and density $n=1$ is indicated by the gray line.
Right panel:  sketch of the paramagnetic state DCA phase  diagram of the Hubbard model, calculated for the cluster shown in the left panel at half filling, as a function of interaction strength $U$ and next-nearest-neighbor hopping $t'$. A Fermi-liquid metal phase (left, red online), a sector-selective intermediate phase (middle, green online) and a fully gapped insulating phase (right, blue online) are shown. 
}
\label{cluster}
\end{figure}

The  cluster dynamical mean-field method \cite{Maier05} provides theoretical access to the momentum dependence of the electron self-energy and spectral function. Application of this method to  the two-dimensional Hubbard model has revealed strong indications of momentum-space differentiation and pseudogap formation\cite{Jarrell01,Parcollet04,Civelli05,Kyung06,Macridin06,Chakraborty08,Haule07,Park08plaquette,Gull08_plaquette,Stanescu06,Ferrero09,Ferrero09b,Sakai09,Liebsch09} as well as avoided \cite{Haule07c} or actual \cite{Vidhyadhiraja09} quantum criticality.  In a previous paper\cite{Werner098site} we demonstrated that when applied to an eight-site cluster  the method yields a multistage, momentum-sector-selective  metal-insulator transition in which different regions of the Fermi surface are successively gapped as carrier concentration or interaction strength are varied. The successive gapping bears an intriguing similarity to the behavior of high-$T_c$ cuprates in the pseudogap regime. In this paper we present a detailed analysis of this important phenomenon. We determine the nature of the transitions,  showing, in particular, that the initial transition from a Fermi-liquid like phase to a sector-selective phase corresponds to the continuous opening of a gap (or pseudogap, for a discussion see Sec.~\ref{Summary}) in one momentum sector. The ungapped momentum sector is shown to exhibit non-Fermi-liquid features including a self-energy which vanishes more slowly than linearly as the frequency tends to zero. We determine how the behavior changes with carrier concentration, interaction strength, and variations in band structure, clarify  the effect of breaking particle-hole symmetry and show that the Van Hove singularity does not cause the effects. We also present data indicating that the  phenomenon is linked to a magnetic instability and that the model is not unstable to nematic order.

Most previous cluster dynamical mean-field studies have used interpolation methods to infer the behavior of lattice quantities from the limited information provided by the dynamical mean-field methods and analytical continuation methods to infer the frequency dependence. We believe that it is essential to understand the behavior of the model directly; we therefore avoid ``periodization'' and continuation methods in this paper, basing our conclusions  on the analysis of directly measured quantities.

As first noted by Biermann {\it et al.} \cite{Biermann05b}  the phenomenon of momentum-space differentiation is related to the orbitally selective Mott transition\cite{Anisimov02} which is found in the single-site dynamical mean-field theory of models with orbital degeneracy and has  been extensively studied in the context of transition-metal oxides and actinides with degenerate $d$ or $f$-levels as models for local-moment formation\cite{Koga04,Liebsch05,Biermann05b} and as possible explanations of certain aspects of heavy fermion behavior.\cite{Pepin08}  The connection was explored  by Ferrero {\it et al.} \cite{Ferrero09} in a study of a two-site dynamical mean-field approximation with a cleverly chosen cluster geometry and was also adopted by Liebsch {\it et al.} \cite{Liebsch08} In this paper we attempt to relate our results to the general understanding of the orbital-selective Mott phenomenon and make some comparison to the previous work of Ferrero {\it et al.} and Liebsch {\it et al.}   

An important question which we are not fully able to resolve is the relation between the sector-selective transition we find and the onset of long-ranged order. We find an association between the gap and a pinning of density at a commensurate value which leads us to identify the transitions as Mott transitions. Our calculations are carried out in the paramagnetic phase of the model, where by construction no long-ranged order is possible. It is however quite conceivable that the onset of some sort of long-ranged order underlies the behavior we study.  In four-site clusters this is believed to be the case: the gapping effect was related to the onset of short-ranged order of the plaquette singlet type in Ref.~\onlinecite{Gull08_plaquette} while in Ref.~\onlinecite{Macridin08} a susceptibility analysis was used to argue that a transition to a state with long-ranged columnar dimer order  dominated the physics of the four-site cluster. We present evidence here that antiferromagnetic correlations with range larger than the cluster size are important and we note in passing that nematic order appears not to be favored in the model we study.

The rest of this paper is organized as follows. In Sec.~\ref{Model} we define the model and crucial parameters and outline the theoretical methods. In Sec.~\ref{Analysis} we explain how the phase boundaries are determined and in Sec.~\ref{Phase Diagram} we present the calculated phase diagram. In Secs.~\ref{Characterization},  \ref{Spins}, and \ref{Pomeranchuk} we discuss the physical content of our results. In Sec.~\ref{Selective} we place our results in the context of other work on orbital-selective transitions. Section \ref{Summary} is a conclusion, and is organized so that readers  may turn directly to this section,  which summarizes the results with pointers back to appropriate portions of the main text.

\section{Model and method \label{Model}}

We study  the two-dimensional one-band Hubbard model with Hamiltonian
\begin{equation}
H = \sum_{p,\sigma} \left(\epsilon_{p} -\mu\right)
c^\dagger_{p,\sigma}c_{p,\sigma}+U\sum_i n_{i,\uparrow}n_{i,\downarrow},
\label{H}
\end{equation} 
where
\begin{equation}
\epsilon_p=-2t\left[\cos(p_x)+\cos(p_y)\right]-4t'\cos(p_x)\cos(p_y).
\label{dispersion}
\end{equation}

The low ($\omega \lesssim 4\text{eV}$)-energy physics of the high-temperature superconductors is believed to be described by a model of this sort with interaction  $U \sim 9t$ and next-nearest-neighbor hopping  $-0.3t \lesssim t' \lesssim -0.15t$.  A non-vanishing $t'$ breaks particle-hole symmetry and we shall see that this has an important effect on the results. The carrier concentration $n$ is controlled by the chemical potential $\mu$ and we shall be interested in dopings $x=1-n$ between $\pm 0.3$.

We use the ``Dynamical Cluster Approximation'' (DCA) formulation of cluster dynamical mean-field theory (Refs.~\onlinecite{Hettler98} and \onlinecite{Maier05}) in which the Brillouin-zone is divided  into  $N$  ``patches'' defined by the basis functions $\phi_\alpha(p)$ which are 1 for $p$ in patch $\alpha$ and zero otherwise and the electron self-energy is approximated as
\begin{equation}
\Sigma(p,\omega) \rightarrow \sum_{\alpha=1}^{N}\phi_\alpha(p)\Sigma_\alpha(\omega).
\label{sigmadca}
\end{equation}

The patches are required to have equal area and to fully tile the Brillouin-zone. Different DCA approximations are defined by the number of patches and the choice of patch shapes and positions. We  study here the $N=8$-site DCA approximation defined by the tiling shown in  Fig.~\ref{cluster}.  We label the patches by the $8$ momenta $K_{i=1...8}$ at the patch centers and distinguish inequivalent momentum sectors by letters $A$, $B$, $C$, and $D$ as shown. An advantage of this tiling is that for the carrier concentrations of interest the Fermi surface of the non-interacting model passes through  two inequivalent sectors (labeled $B$ and $C$). The cluster therefore gives direct (if coarse grained) access to momentum variation around the Fermi surface. In previous studies of two-site \cite{Ferrero09} and four-site \cite{Parcollet04,Civelli05,Kyung06,Macridin06,Chakraborty08,Haule07,Park08plaquette,Gull08_plaquette,Stanescu06,Liebsch09} clusters, variations around the Fermi surface had to be inferred from data involving also the momentum sectors $(0,0)$ and $(\pi,\pi),$ which are far from the Fermi surface so that variation away from the Fermi surface was mixed in with variation around the Fermi surface.

The frequency-dependent functions $\Sigma_\alpha(\omega)$, $\alpha=1,\ldots,8$ are obtained from the solution of an eight-site quantum impurity model\cite{Georges96,Maier05} defined by an action $S=\int d\tau d\tau'{\cal L}(\tau,\tau')$ which is a function of a continuous (imaginary)-time variable and the momentum sector labels ${K}_{i=1...8}$ corresponding to the centers of the tiles shown in Fig.~\ref{cluster}. Specifically,
\begin{align}
{\cal L}&=\sum_{K}{\cal G}^{-1}_K(\tau-\tau')d^\dagger_{K,\sigma}(\tau)d_{K,\sigma}(\tau')\\ \nonumber&+U\sum^{'}_{K_1...K_4}d^\dagger_{K_1,\sigma}(\tau)d_{K_2,\sigma}(\tau)d^\dagger_{K_3,\sigma'}(\tau)d_{K_4,\sigma'}(\tau)
\label{Limp}
\end{align}
with $\sum^{'}$ denoting a sum over all $K$ such that $K_1+K_2+K_3+K_4$ equals a reciprocal-lattice vector. The mean-field functions ${\cal G}_K^{-1}$ are obtained from 
\begin{equation}
{\cal G}_K^{-1}=\Sigma_K+\left[\int _K(dk)G(k)\right]^{-1}
\label{dcasce}
\end{equation}
with $\int_K (dk)$ standing for an integral over the tile centered on momentum $K$, normalized so that $\int_K (dk) 1=1$ and  $G(k)$ is the Green's function of the lattice problem computed with $\Sigma$ defined by Eq.~(\ref{sigmadca}). Note that within the DCA approximation the different momentum sectors are not mixed by the self-consistency condition. 

Following the suggestion of Ref.~\onlinecite{Biermann05b}, Ferrero {\it et al.} \cite{Ferrero09} and Liebsch {\it et al.} \cite{Liebsch08} observed that  the DCA equations have the same form as those used in the single-site dynamical mean-field theory of a multiorbital system such as a transition-metal oxide or the $f$ levels in a heavy fermion material. In the latter case one identifies the labels $K$ with the different local orbital states, the $\int _K$ becomes the integral over the whole zone of an appropriate projection of a multiband lattice Green's function, and the self-consistency equation is in general not diagonal in the indices $K$.

We solve the impurity model with the numerically exact continuous-time auxiliary field technique \cite{Gull08_ctaux} with delayed updates.\cite{Alvarez08}  Calculations are rendered difficult by a fermion  sign problem which causes the computation to scale exponentially instead of polynomially in inverse  temperature. The sign problem vanishes in the particle-hole symmetric case  $t'=0$,  $n=1$ but becomes more severe as $t'$ and $|n-1|$ are increased. The severity of the problem may be characterized by the average sign $\langle s\rangle$ associated with the simulation of a fermionic problem (see, e.g., Ref.~\onlinecite{Troyer05}) because errors scale roughly as $1/\langle s\rangle$. In sign-problem-free situations $\langle s\rangle=1$ at all temperatures but in the typical cases studied here $\langle s \rangle<1$ and decreases exponentially as temperature $T\rightarrow 0$. The  severity of the sign problem increases with $U$ and $t'$ and is apparently more severe for moderate than for small or large doping.  For example,  at interaction strength $U=7t$, $t'/t = -0.3$ and inverse temperature $\beta t=25$ we find $\langle s\rangle=0.15$ for $\mu/t = 1$ (around 5\% hole doping, near the boundary of the sector-selective transition) but $\langle s\rangle \simeq 0.85$ at $U/t = 8,$ $\mu/t = 2$  far from the boundary. The sign problem is also strongly dependent on the cluster geometry.

The lowest temperature accessible with the computational resources available to us is $\beta t\approx 40$ at $U < 8t$ and most of our results are obtained at $\beta t\sim 20$.   The conventional high-$T_c$ band parametrization\cite{Andersen95} implies that $\beta t=20$ corresponds to temperature $T\sim 200\text{K}$.

\begin{figure}[htb]
\begin{center}
\includegraphics[width=0.95\columnwidth]{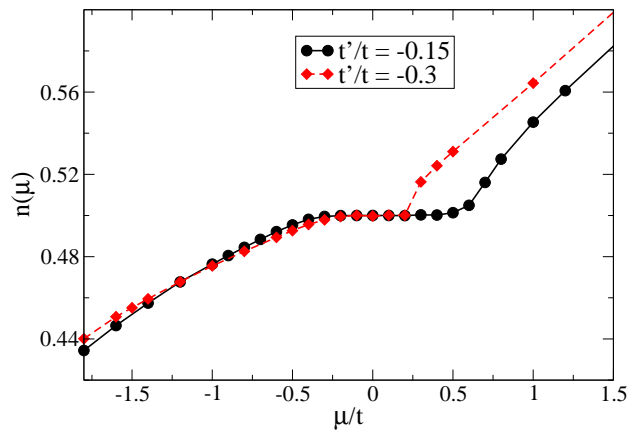}\\
\end{center}
\caption{Total density per spin as a function of chemical potential, for $U/t = 7$, $\beta t=20$, and second-neighbor hopping $t'/t = -0.15$ (solid line and circles, black online) and $t'/t=-0.3$ (dashed line and diamonds, red online)}
\label{ntotvsmu.fig}
\end{figure}
Throughout the paper we mainly present results as a function of chemical potential. For reference, we show in Fig.~\ref{ntotvsmu.fig} the total density as a function of chemical potential, for $U/t = 7$, $\beta t=20$, and  two representative values of the next-neighbor hopping $t'$.
\begin{figure}[htb]
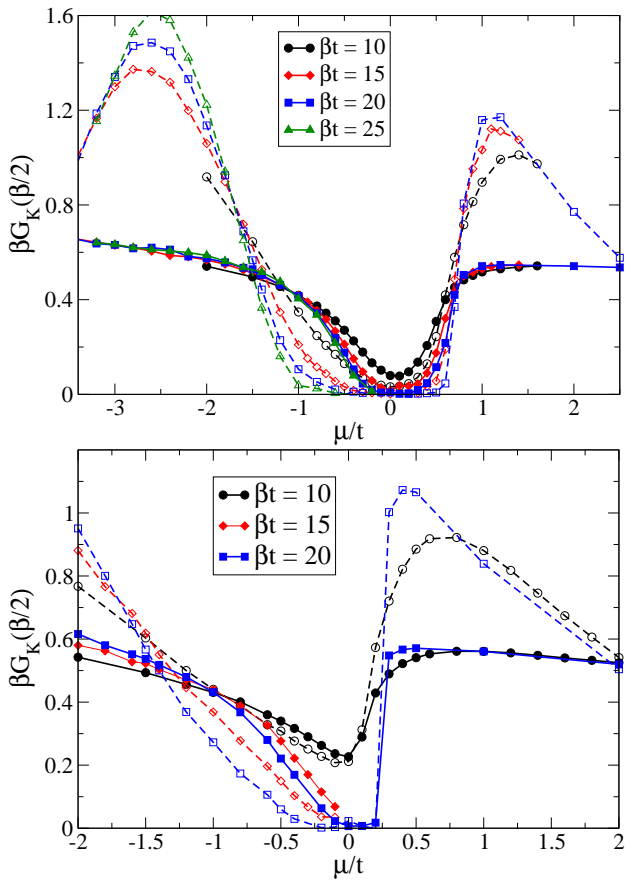

\begin{center}
\includegraphics[width=0.95\columnwidth]{ghalf_mu_tprime0.15.eps}\\
\includegraphics[width=0.95\columnwidth]{ghalf_mu_tprime0.3.eps}\\
\end{center}
\caption{Upper panel: $\beta G(\frac{\beta}{2})$ for sectors $B$ (full symbols) and $C$ (empty symbols), at $U/t = 7$ and $t'/t=-0.15$. The strong temperature dependence in the sector $C$ curves arises from the Van Hove divergence in the density of states. The crossing points indicate the onset of gapping in the sectors.
Lower panel: same, for $t'/t = -0.3.$
}\label{ghalf_mu_tprime0.15.fig}
\end{figure}

\section{Analysis \label{Analysis}}
As noted above we base the conclusions of this paper on the analysis of directly measured quantities. The most important and easily accessible of these are the imaginary-time sector Green's functions 
%\textcolor{red}{EG: I want G to be positive, not negative. We need to change the minus signs here for it to be positive, instead of changing them in the text/figures}
$G_K\equiv \left<T_\tau c_K(\tau)c^\dagger_K(0)\right>=\left({\cal G}_K^{-1}-\Sigma_K\right)^{-1}$,  from which we also may extract the sector density as
\begin{equation}
n_K=-G_K(\tau \rightarrow 0^-).
\label{nkdef}
\end{equation}
Note that the total particle density is $n=\sum_K n_K/8$.

It is useful to consider the value of the sector Green's function at the center of the imaginary-time interval, $G_K(\tau=\frac{\beta}{2})$.  Use of Eq.~(\ref{dcasce}) and the usual spectral representation formulas shows that 
\begin{equation}
\beta G_K\left(\frac{\beta}{2}\right)=\int \frac{d\omega}{2\pi T}\frac{A_K(\omega)}{\cosh[\omega/(2T)]}
\label{gtau}
\end{equation}
with sector spectral function
\begin{equation}
A_K(\omega)=\int _K(dk)\frac{\Im\Sigma_K(\omega)}{(\omega+\mu-\varepsilon_k-\text{Re}\Sigma_K(\omega))^2+\Im \Sigma^2_K(\omega)}.
\label{AK}
\end{equation}

The integral in Eq.~(\ref{gtau}) is dominated by $\omega \lesssim T$  so that in the low-temperature limit the measurement provides information about the Fermi-level sector-averaged electron spectral function. We may distinguish three cases. 

In a Fermi-liquid metal, $\Im \Sigma(\omega\rightarrow 0)$ is negligible and $\mu-\varepsilon_k-\text{Re}\Sigma_K(\omega=0)$ vanishes for some $k$ in the sector. In this case $\beta G_K(\beta/2)$ becomes the non-interacting density of states $A_\text{non}=\pi \int _K(dk)\delta(\mu_\text{eff}-\varepsilon_k)$ evaluated at the renormalized chemical potential
\begin{equation}
\mu_K^\text{eff}=\mu-\text{Re} \Sigma_K(\omega=0).
\label{mustar}
\end{equation}
One important caveat is that sector $C$ contains the Van Hove singularity of the two-dimensional band structure, where the density of states diverges. This divergence leads to a strong temperature dependence in $\beta G(\beta/2)$ so that the non-interacting density of states value may be reached only at very low temperatures, often below our measurement limit. 

In a non-Fermi-liquid metal $\Im \Sigma(\omega)$ vanishes less rapidly than $\omega$ and $\mu=\varepsilon_k-\text{Re}\Sigma_K(\omega=0)$ is satisfied for some $k$ in the sector and as $T\rightarrow 0$ $\beta G_K(\beta/2)$ takes a value which is nonvanishing but typically  less than $A_\text{non}(\mu_\text{eff})$.

Finally, we may have a gapped state if $\Im \Sigma(\omega\rightarrow 0)\rightarrow 0$ and $\mu-\varepsilon_k-\text{Re}\Sigma_K(\omega=0)$ does not vanish for any $k$ in the sector so that $\beta G_K(\beta/2)$ vanishes exponentially at low $T$.  We shall see that in our calculations this effect arises from the appearance of a low-frequency pole in the self-energy which pushes states away from the Fermi level, making it difficult to satisfy the quasiparticle equation at low energies. 

The boundaries of the gapped regions may, in principle, be determined from any of  the density, the Green's function, or the self-energy. In practice significant uncertainties, compounded by our limited temperature range, exist in each method. We believe that the most reliable method is  to plot $\beta G(\beta/2)$ as a function of interaction strength or chemical potential for several temperatures as shown in Fig.~\ref{ghalf_mu_tprime0.15.fig}.  

For sector $C$ on the negative $\mu$ (hole-doped) side a clear crossing point at $\mu\approx-1.5t$  is evident, separating a regime where $\beta G(\beta/2)$ increases as $T$ is decreased from a region where $\beta G(\beta/2)$ decreases as $T$ is decreased. We interpret the crossing point as marking the chemical potential at which a gap begins to open in sector $C$. We observe that for $\mu \lesssim -1.5t$,  $\beta G(\beta/2)$ is substantially less than the Fermi-liquid value but increases as $T$ is decreased. This behavior is consistent with the hypothesis of a Fermi liquid state with very large thermal corrections due to the divergence in the density of states at the Van Hove point. At the temperatures available to us the possibility of a marginal Fermi-liquid or non-Fermi-liquid state in this parameter regime as proposed in Ref.~\onlinecite{Vidhyadhiraja09}  can neither be ruled out nor confirmed.

In sector $B$  the identification of the transition point in the doping case is more complicated (in the interaction-driven case an analysis as in sector $C$ is straightforward). As can be seen from Fig.~\ref{ghalf_mu_tprime0.15.fig} (see also Fig.~\ref{vanhove.fig}) a simple crossing point does not occur in the $\beta G(\beta/2)$ graph. Rather, one begins to see a fan out from a temperature-independent set of curves. Furthermore, in the range $-1.3t<\mu<-0.9t$ $\beta G(\beta/2)$ seems to evolve (see also Fig.~\ref{vanhove.fig}) as $T$ is decreased to a value which is temperature independent but less than the Fermi-liquid value $\beta G_0(\beta/2,\mu_\text{eff})$. This is evidence for a non-Fermi-liquid state. As the chemical potential is further increased, evidence of a temperature-dependent decrease in $\beta G(\beta/2)$ becomes apparent, and by $\mu=-0.5t$ a clear gap has opened.  Precisely locating the point at which the physics changes from a gapless non-Fermi-liquid to a gapped state is thus challenging but it is clear that the onset of gapped behavior in sector $B$ occurs at a substantially higher chemical potential than the onset of such behavior in sector $C$. 

\begin{figure}[b]
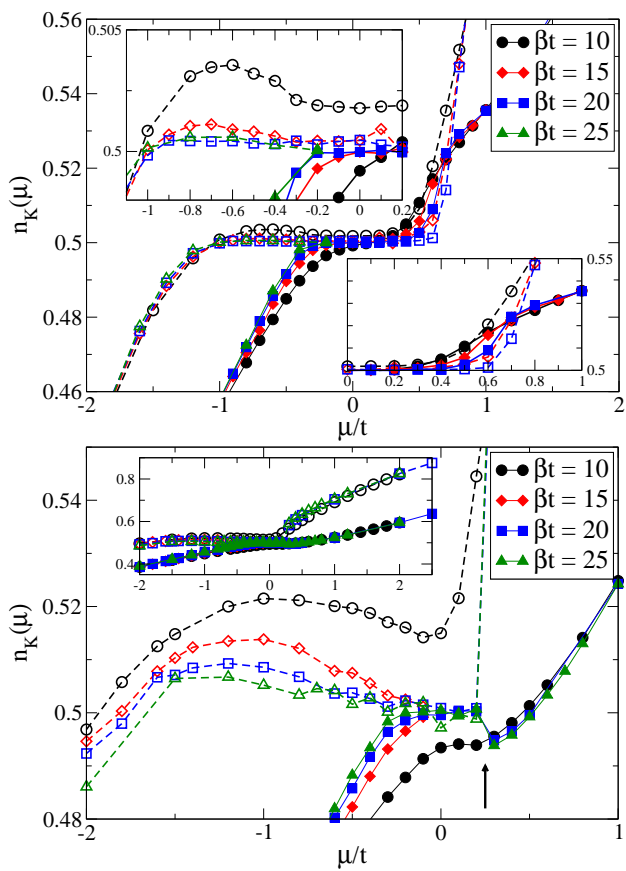

\begin{center}
\includegraphics[width=0.95\columnwidth]{densities_mu_tprime0.15.eps}\\
\includegraphics[width=0.95\columnwidth]{densities_mu_tprime0.3.eps}
\caption{Upper panel: Sector-specific density $n(\mu)$ for $U/t = 7$, $\beta t= 10, 15, 20, 25$, and $t'/t = -0.15$. Open symbols: Sector $C$. Filled symbols: sector $B$.  Insets: expanded view of the hole-doped (upper left) and electron-doped (lower right) metal insulator transition regions. Lower panel: same for $t'/t = -0.3$. The pinning of the density to $1/2$ in the insulating phase is clearly visible. Arrow: location of the first-order transition.}
\label{densities_mu_tprime0.15.fig}
\end{center}
\end{figure}

One may also consider density vs chemical-potential curves such as those shown in Fig.~\ref{densities_mu_tprime0.15.fig}.  Interpretation  is complicated by two issues: first, if the gap is small then, at the temperatures accessible to us, a substantial temperature variation occurs. Second, while our data indicate that if sector $K$ has a gap, then at $T=0$ $n_K = 1/2$, we see that the density may approach the pinned value from above or below depending on the value of the chemical potential. For these reasons we do not use the $n_K$ data to identify phase boundaries.

Alternatively, the phase boundaries can be determined by the evolution of the self-energy or inverse self-energy with temperature. We chose not to use this method, as self-energies in the sector-selective region display a strong temperature dependence that can be analyzed more accurately by considering the crossing point of $\beta G(\beta/2)$.

\begin{figure}[t]
\begin{center}
\includegraphics[angle=270,width=0.95\columnwidth]{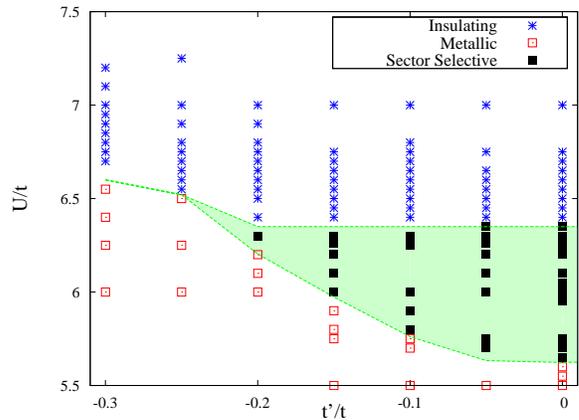}
\caption{Phase diagram calculated at half filling in the plane of interaction strength $U$ and second-neighbor hopping $t'$  from measurements of 
the crossing point of $\beta G(\beta/2)$ at $\beta t = 15$ and $20$ as described in Sec.~\ref{Analysis}. 
Lower region (open squares, red online): ``metallic'' region with gapless behavior in sectors $B$ and $C$. 
Upper region (stars, blue online), ``insulating'' region with all sectors gapped. Shaded area (filled squares, black online): 
sector-selective region with sector $C$ gapped and sector $B$ gapless. }
\label{PDHalffilled}
\end{center}
\end{figure}

\section{Phase Diagram\label{Phase Diagram}}
\subsection{Interaction-driven transition, half filling}
Figure~\ref{PDHalffilled} shows the phase diagram determined at half filling as a function of interaction strength $U$ and second-neighbor hopping $t'$ (note that at $n=1$ the phase diagram depends only on $|t'|$). For $t'$ not too large, a two-stage transition occurs, in which as the interaction is increased a gap first opens in sector $C$ followed by a second transition to a completely gapped state.   However, as the second-neighbor hopping amplitude is increased, the range in $U$ over which only one sector is gapped decreases and for $|t'| \gtrsim 0.25t$ there is only one transition.

\begin{figure}[b]
\begin{center}
\includegraphics[width=0.95\columnwidth]{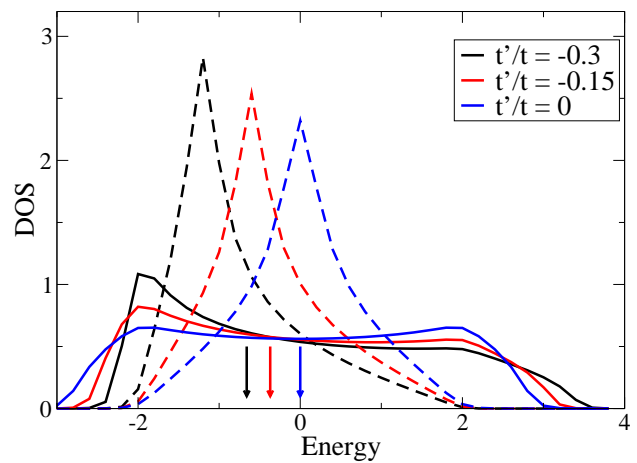}
\caption{Noninteracting density of states computed from the analytically known $\beta G_0(\beta/2,\mu)$ at inverse temperature $\beta=25t$ for $t'/t =0, -0.15$, and $-0.3$. Bold lines: sector $B$. Dashed lines: sector $C$. Sectors $A$ and $D$ are not shown. Arrows indicate the chemical potential corresponding to half filling.
}
\label{nonint_diff.fig}
\end{center}
\end{figure}

We may relate the orbitally selective behavior to the density of states ${\cal D}(\varepsilon)=\int (dk)\pi \delta(\varepsilon-\varepsilon_k)$ shown for the two relevant sectors in Fig.~\ref{nonint_diff.fig}. At $t'=0$ the densities of states of sectors $B$ and $C$ are symmetrical about $\varepsilon=0$ (the chemical potential corresponding to half filling) but the density of states in sector $C$ is sharply peaked at $\varepsilon$ corresponding to the Van Hove energy $\varepsilon(p_x=0,p_y=\pi/2)$ ($=0$ at $t'=0$) and is narrower than that of sector $B$.  It is natural that the sector with the narrower band and the higher density of states should undergo the Mott transition first.\cite{Anisimov02,Koga04,Biermann05b,Koga05,Jakobi09} As $t'$ is changed from zero the position of the Van Hove point shifts in energy, although the positions of the band edges of sector $C$ do not change [the bandwidth of sector $B$ does change from $4\sqrt{2}t - 4t'$ to $4t$ for $|t'| > (\sqrt{2}-1)t$]. The chemical potential corresponding to half filling similarly shifts.  The existing literature on the orbitally selective Mott phenomenon does not provide guidance on the fate of the sector-selective transition in these circumstances. We note, however, that a metric for estimating the location of the Mott transition is to compare the interaction energy to the $U=0$ kinetic energy given by $\int_{sector}(\varepsilon_p-\mu)f(\varepsilon_p-\mu)$ with $f$ the Fermi function.  As $t'$ is shifted from zero to $-0.3t$ the kinetic energies of the $B$ and $C$ sectors change from $(-0.084,-0.018)$ to $(-0.035,-0.023)$; thus as $t'$ is changed from zero the kinetic energies of the bands become similar, explaining the disappearance of the orbitally selective phase.  

\subsection{Doping-driven transition, fixed interaction}

\begin{figure}[b]
\begin{center}
\includegraphics[width=0.95\columnwidth]{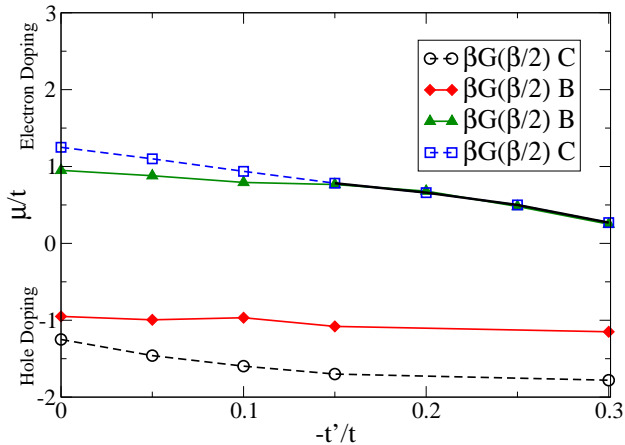}
\caption{Phase diagram at  interaction strength $U=7t$ as a function of chemical potential $\mu/t$ and second-neighbor hopping  $t'/t$, from measurements of  $\beta G(\beta/2)$. Positive $\mu$ corresponds to electron doping and negative $\mu$ to hole doping.  A particle-hole transformation converts $t'\rightarrow -t'$ so it is sufficient to show  only one sign of  $t'$  but both polarities of doping. Dashed lines (circles and squares) indicate chemical potentials at which the $C$-sector gap opens, heavy solid lines (diamonds and  triangles) indicate chemical potentials where the $B$-sector gap opens.  Note that for $-0.15t$ the critical chemical potential $\mu_\text{eff}$ at which the sector-selective transition occurs corresponds to a hole doping of $x \approx 0.109$. Thick black line: location of the coalesced first-order transition.
}
\label{PDDoping}
\end{center}
\end{figure}

Sector-selective transitions occur also as functions of doping at fixed interaction.  For interaction strengths lying in the orbitally selective (shaded)  region of Fig.~\ref{PDHalffilled} sector $B$ remains  gapless as the chemical potential is changed whereas there is a critical doping level (or chemical potential) beyond which the sector $C$ gap closes. We do not consider this case further, focusing instead on $U$ large enough that  both sectors are gapped at half filling.   Figure~\ref{PDDoping} presents a phase diagram in the space of chemical potential and $t'$ for  $U=7t$.  For small $|t'|$, both electron and hole doping are seen to produce a two-step transition, in which the first stages of doping take place in sector $B$ and only subsequently does sector $C$ start to be doped. Interestingly, as $-t'$ is increased a qualitative particle-hole asymmetry develops. On the electron-doped side the two transitions coalesce (within our resolution) to one and become first order\cite{Macridin06b,Liebsch09} while on the hole-doped side the presence of  two transitions appears to be a generic feature. 

\section{Characterization of  transitions\label{Characterization}}
\subsection{Interaction-driven transition}
In this section we  characterize the interaction-driven transitions found in the eight-site cluster. We begin by summarizing what is known about metal-insulator transitions in other dynamical mean-field theory  implementations.  In the single-site dynamical mean-field approximation the correlation-driven metal-insulator transition occurs at a large $U$ and has a complicated structure in which a preformed, large-magnitude gap exists on both sides of the transition and metallic vs insulating behavior is determined by the presence or absence of midgap states.\cite{Georges96} On the other hand, in four-site cluster DMFT approximations the transition occurs at a relatively small $U\sim 5t$ and is apparently always first order, characterized by the discontinuous opening of a gap as the correlation strength is increased above a critical value.\cite{Gull08_plaquette,Park08plaquette}

\begin{figure}[tbh]
\begin{center}
\includegraphics[width=0.95\columnwidth]{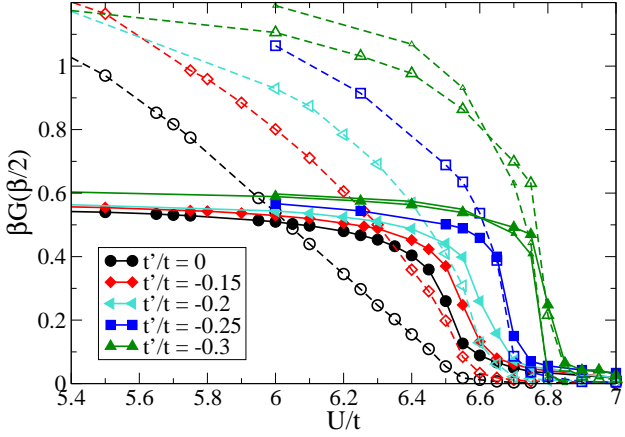}
\caption{$\beta G(\beta/2)$ at density $n=1$ for sector $B$ (filled symbols) and $C$ (open symbols) as function of interaction strength $U/t$ for several values of the next-nearest-neighbor hopping $t'/t$ at $\beta t=15$. At $t'/t = -0.3$ $\beta t=20$ data are also shown (thin line, small symbols).}
\label{ghalf_interaction.fig}
\end{center}
\end{figure}

Figure~\ref{ghalf_interaction.fig} shows our results for the sector density of states for several $t'$ values. We observe, in agreement with Fig.~2 of Ref.~\onlinecite{Werner098site} which presents data for only two $t'$ but at the lower temperature $\beta t=40$, that for $t{'}$ near zero the two transitions are well separated in $U$ and that, while the $C$-sector transition appears smooth, marked by a crossing point in $\beta G(\beta/2)$, the $B$-sector transition seems discontinuous, marked by an apparent jump in $G(\beta/2)$. As the magnitude of $t{'}$ increases, the $C$ transition steepens, and beyond the point where the sector-selective phase disappears the single transition becomes first order as in the four-site case. 

We may understand the change in transition order from consideration of the sector densities. The breaking of particle-hole symmetry caused by a nonzero $t'$ means that in the weakly correlated metallic phase at total density $n=1$ $n_C \neq 1/2.$ For $t'$ small, $n_C$ is close to $1/2$ and evolves smoothly to the pinned value as $U$ approaches the sector-selective value. However for larger $t'$ the difference of $n_C$ from $1/2$ is too large, and the evolution with $U$ is preempted by a first-order transition.

\begin{figure}[htb]
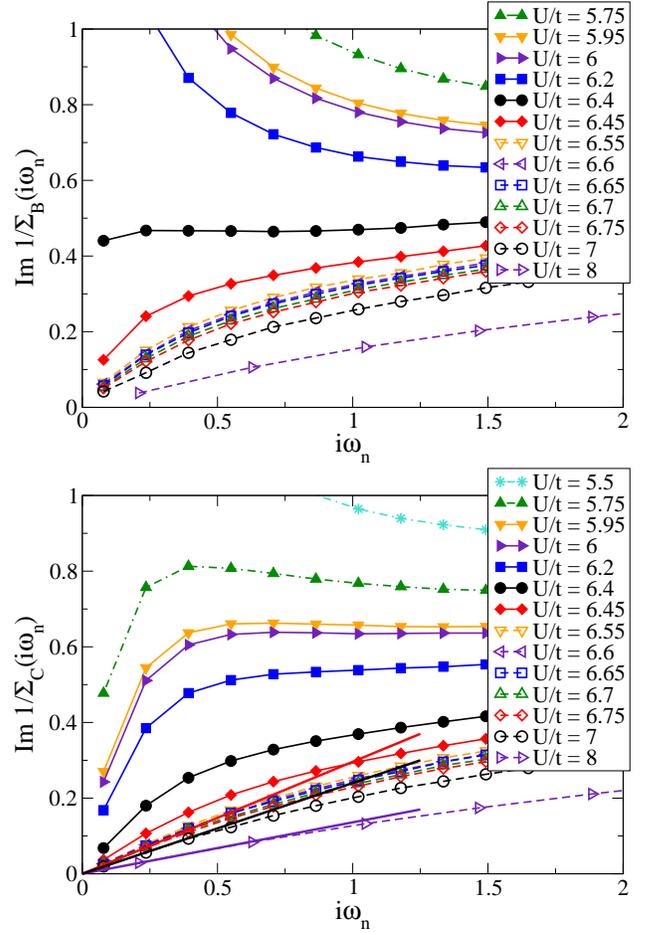

\begin{center}
\includegraphics[width=0.95\columnwidth]{ImSelfenergiesInvUSectorB_tprime0_lowT.eps}\\
\includegraphics[width=0.95\columnwidth]{ImSelfenergiesInvUSectorC_tprime0_lowT.eps}
\caption{Imaginary part of the inverse of the Matsubara-axis self-energy for momentum  sectors $B$ and $C$, measured as a function of interaction strength $U/t$ at density $n=1,$ $t'/t = 0$ and inverse temperature $\beta t=40$ ($\beta t=15$ for $U/t = 8$). Solid lines indicate fits to $U/t = 6.75, 7,$ and $8$ with slopes $\Im \Sigma_C^{-1}(\omega_0)/\omega_0 = 0.14, 0.25,$ and $0.30$ (in units of $t^{-2}$), respectively.
}
\label{inv_interact_0.fig}
\end{center}
\end{figure}

We now show how this behavior is reflected in the self-energy. We expect the opening of a gap to be related to the appearance of a pole in the self-energy so that 
\begin{equation}
\Im \Sigma_K(\omega)=\pi \Delta^2 \delta(\omega-\omega_P^K)+\Sigma^{''}_{reg}(\omega)
\label{sigpole}
\end{equation}
implying
\begin{equation}
\Sigma_K(i\omega_n)=\frac{\Delta^2}{i\omega_n-\omega^K_P}+A_0+A_1i\omega_n+\cdots,
\label{sigmat}
\end{equation}
where $A_{0,1}$ and the ellipsis arise from the regular part $\Sigma_{reg}$ of $\text{Im} \Sigma$.  Reference \onlinecite{Wang09a} found that correlation gaps in Hubbard-like models were reliably estimated from the solution to the quasiparticle equation $\omega-\varepsilon_k+\mu-\text{Re}\Sigma(\omega)=0$. If we neglect the $A$ terms we may solve the quasiparticle equation to obtain an estimate for the sector half gap $E^K_g$ in terms of the band-edge energy $E^K=2\sqrt{2}t-2t'$ [sector $B$, $|t'|<(\sqrt{2}-1)t$] and $2$ (sector $C$),

\begin{equation}
E_g\approx \frac{\sqrt{(E^K+\omega_p)^2+4\Delta^2}+\sqrt{(E^K-\omega_p)^2+4\Delta^2}-2E^K}{2}.
\label{gap}
\end{equation}

In the particle-hole symmetric case $\omega_P=A_0=0$,  so $\Sigma(i\omega_n)$ is purely imaginary and would diverge as $\omega_n\rightarrow 0$ while  $E_g=\sqrt{(E^K)^2+4\Delta^2}-E^K.$ Figure~\ref{inv_interact_0.fig} reveals essentially this behavior. In sector $C$ at $U=7t$ drawing a straight line through the origin at the lowest Matsubara point yields $\Delta^2\approx 4/t^2,$ implying $E_g\approx 1.25 t$, reasonably consistent with the gap defined from the chemical potential in Fig.~\ref{PDDoping} while in sector $B$ a similar analysis gives (with less confidence) $0.7t$, slightly smaller than the chemical-potential gap. 

As the interaction strength is decreased from $U=7t$ to $U=6.55t$ the $C$- and $B$-axis self-energies and therefore gaps change only slowly. However, as the interaction is further decreased from $6.55t$ to $6.45t$ to $6.4t$ the evidence for a gap in sector $B$ vanishes abruptly and the gap magnitude in sector $C$ becomes smaller. As the interaction strength is yet further decreased the $C$-sector gap drops rapidly. The frequency range implied by the temperatures shown does not permit a reliable gap analysis, and, in particular, does not exclude the possibility of a pseudogap such as discussed in Ref.~\onlinecite{Ferrero09b} but the rough estimate obtained by drawing a straight line from the origin through the lowest Matsubara frequency indicates that the gap (or pseudogap) at $U=6.2t$ is about a factor of four smaller than that at $U=6.65t$.
%Finite temperature effects in the self-energy are relatively strong. %and hamper a similar analysis away from particle hole symmetry.
%\textcolor{red}{Maybe say something about T-effects and why the phase boundary of fig. 4 is apparently shifted compared to Fig. 8}

\begin{figure}[htb]
\begin{center}
\includegraphics[width=0.95\columnwidth]{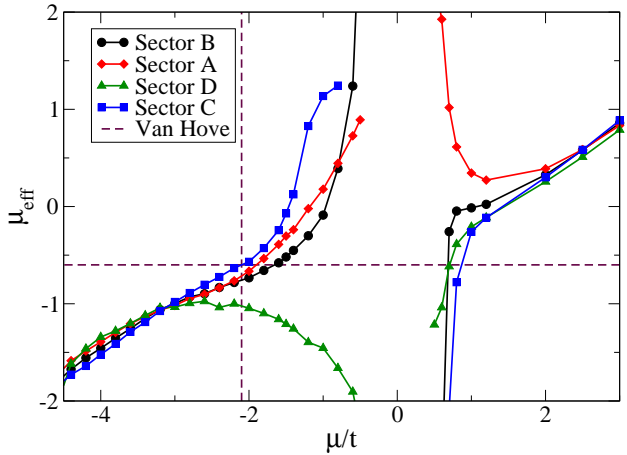}
\caption{Sector-dependent effective chemical potential $\mu_\text{eff}=\mu - \text{Re}\Sigma(\omega=0)$, plotted as a function of chemical potential for $U/t = 7, t'/t = -0.15$, and $\beta t=20$. The horizontal dashed line indicates the $\mu_\text{eff}$ value corresponding to the Van Hove point $\mu_\text{VH}=-0.6$ in the $C$-sector density of states (Fig.~\ref{nonint_diff.fig}) and the vertical dashed line indicates the $\mu$ value at which $\mu_{\text{eff},C} = \mu_\text{VH}.$ The ``effective'' location of the Van Hove singularity in sector $C$ is shifted to $\mu/t = -2.1$. Values in the insulating region are excluded. $\text{Re}\Sigma$ was obtained by linearly extrapolating the first two frequencies.}
\label{mueff.fig}
\end{center}
\end{figure}

\subsection{Doping-driven transition}
In this section we characterize the doping-driven transition at a correlation strength sufficiently strong that both sectors are insulating at $n=1$.  We begin with an analysis of the dependence of density on chemical potential, shown in Fig.~\ref{densities_mu_tprime0.15.fig}.  The first noteworthy point is that on the hole-doping side the sector $C$ transition occurs at the chemical potential corresponding to sector $C$ being half filled.  The upper left inset shows an expanded view of the $C$-sector occupancy for different temperatures, demonstrating that deviations from the half-filled value are due to thermal excitations in a particle-hole asymmetric situation. We have found similar results for all other parameters  and for this reason identify the sector-$C$ transition as an orbitally selective Mott transition.

The important role played by the particle density is further demonstrated by the lower right inset of the upper panel of Fig.~\ref{densities_mu_tprime0.15.fig},
which provides an expanded view of the electron-doped transition. We see that as $T\rightarrow 0$ the $C$-sector occupancy apparently develops a discontinuous jump. We observe that in the non-interacting model at  $t'<0$, half filling corresponds to the $C$ sector being more than half filled and the $B$ sector being less than half filled, with the filling  difference increasing as the magnitude of  $t'$ increases.  Thus as we approach half filling from the electron-doped side, encountering a  sector-selective transition means shifting electrons from sector $C$ to sector $B$.  For $t'$ near $0$ this is possible; for larger  amplitude $t'$ this becomes too energetically expensive. The two transitions coalesce into one first-order transition. This can be seen in the lower panel of Fig.~\ref{densities_mu_tprime0.15.fig} for $t'/t=-0.3$ which shows that the density of the $B$ sector decreases below the value of half filling when the chemical potential is raised (indicated in the figure by an arrow) while the $C$-sector density jumps vertically to around 0.59.

%\textcolor{magenta}{AJM: CAN WE FIND A DOPING WHERE WE CAN IDENTIFY THE POLE POSITION AND SEE IF THE SECTOR C POLE IS IN THE MIDDLE OF ITS GAP. NEED TO THINK ABOUT THIS.} 

\begin{figure}[htb]
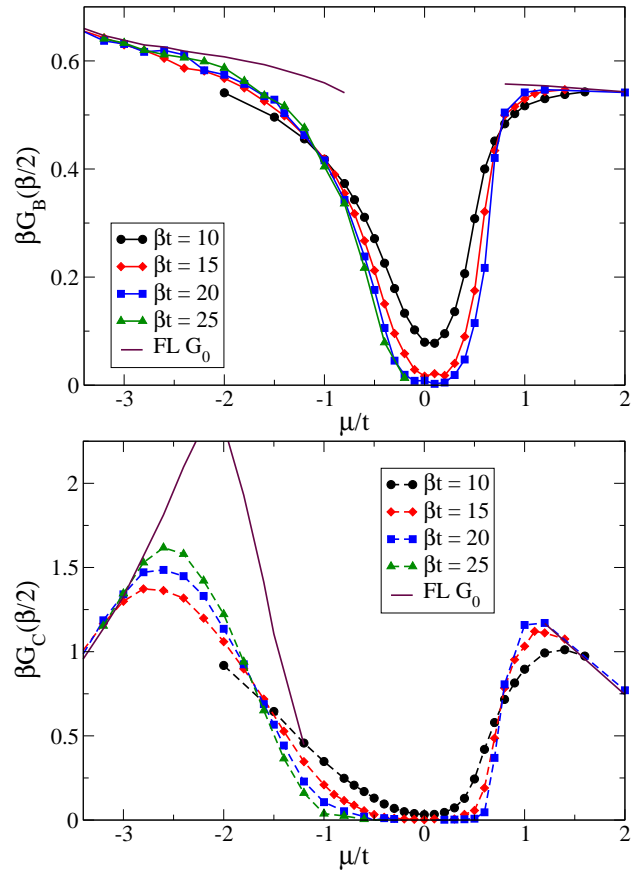

\begin{center}
\includegraphics[width=0.95\columnwidth]{g0mueffB.eps}
\includegraphics[width=0.95\columnwidth]{g0mueffC.eps}
\caption{
Zoom of Fig.~\ref{ghalf_mu_tprime0.15.fig}: estimator $\beta G(\beta/2)$ for the many-body density of states calculated as a function of chemical potential $\mu$ for $U/t =7$ and $t'/t =-0.15$ for sector $B$ (upper panel) and sector $C$ (lower panel). Solid line (purple online): Fermi-liquid value $\beta G(\beta/2)$ evaluated for noninteracting electrons at the chemical potential $\mu_\text{eff}$.
}
\label{vanhove.fig}
\end{center}
\end{figure}
An alternative view of this physics is provided by Fig.~\ref{mueff.fig}, which displays the renormalized chemical potential $\mu_\text{eff}$ defined from the zero-frequency limit of the self-energy via Eq.~(\ref{mustar}) for the different sectors and $t'=-0.15t$. For low carrier concentrations $\mu<-3t$  the self-energy is almost momentum independent and the difference between $\mu_\text{eff}$ and $\mu$ simply reflects the correlation-induced renormalization of the $n(\mu)$  curve familiar from elementary analyses. As $\mu$ is increased through $-3t$ momentum-space differentiation begins to occur. Of particular interest  here is that the $C$-sector $\mu_\text{eff}$ lies above the $B$-sector $\mu_\text{eff}$, signaling the onset of a relative shift of carriers from $C$ to $B$. The  dashed lines indicate the renormalized chemical potential at which the $C$-sector Fermi surface passes through the Van Hove point.  Figure~\ref{mueff.fig} shows that the $\mu_\text{eff}(\mu)$ curves pass through $\mu_\text{eff}=\mu_\text{VH}$ with no noticeable change in behavior. Figure~\ref{vanhove.fig} shows that the many-body density of states evolves smoothly as $\mu$ is increased through the point where $\mu_\text{eff}=\mu_\text{VH}$. The solid curve in the lower panel of Fig.~\ref{vanhove.fig} shows $\beta G(\beta/2)$ computed for non-interacting electrons at $\mu=\mu_\text{eff}$. The Van Hove peak is evident and occurs at a significantly lower $\mu$ than the crossing point $\mu_C \simeq -1.5t$, at which sector $C$ becomes gapped. This physics is also signaled by the rapid upturn in $\mu_{\text{eff},C}$, which occurs because for $\mu > \mu_C$ $n_C$ is pinned so the remaining doping is forced into sector $B$. Finally we note that on the electron-doped side the relative ordering of the $\mu_\text{eff}$ is opposite, signaling a transfer of carriers from $C$ to $B$.

The upper panel of Fig.~\ref{vanhove.fig} reveals an additional interesting feature: a range from $\mu \approx -1.5t$ to $\mu \approx -0.7t$, where the estimator for the $B$-sector density of states, $\beta G_B(\beta/2)$ apparently reaches a stable low-$T$ limit which is neither zero nor the Fermi-liquid value. Examination of the $n(\mu)$ plots in Fig.~\ref{densities_mu_tprime0.15.fig} confirms that the sector $B$ transition point is $\approx -0.5t$ so that in the whole non-Fermi-liquid range $-1.5t < \mu < -0.5t$ sector $B$ is not gapped. We interpret this as a non-Fermi-liquid phase for sector $B$, which is approximately coincident with the sector-selective phase. The temperature dependence visible in the sector $C$ plot suggests that for $\mu<-1.5t$ both sectors have a Fermi-liquid ground state, although much lower temperatures would be required to definitively establish this.  Non-Fermi-liquid physics in orbitally selective Mott regimes of models of orbitally degenerate transition-metal oxides was  reported by Refs.~\onlinecite{Biermann05b} and \onlinecite{Liebsch05}. However, our finding of more Fermi-liquid like behavior in the orbitally selective regime of the interaction-driven case suggests that there is not a generic association between sector selectivity and a non-Fermi-liquid behavior of the ungapped sector.

We are aware of three possible physical interpretations for the  apparent non-Fermi-liquid phase indicated by this analysis. It may be intrisically non-Fermi-liquid, indicating that over a nonzero range of the Fermi surface the imaginary part of the electron self-energy vanishes less rapidly than $\omega$ as $\omega\rightarrow 0$. Alternatively, it might be that the gap which is observed in sector $C$ is also present over some of the momentum range included in sector $B$ so that the measured non-Fermi-liquid behavior arises as an average over some parts of the Fermi surface which are gapped and some which are gapless and Fermi-liquid like. Finally, we can of course not rule out the possibility that there is a Fermi-liquid scale which is simply far below our measurement temperatures. Distinguishing these possibilities requires studies of larger clusters and lower temperatures, which are at or beyond the limits of our present computational capabilities. 
\begin{figure}[t]
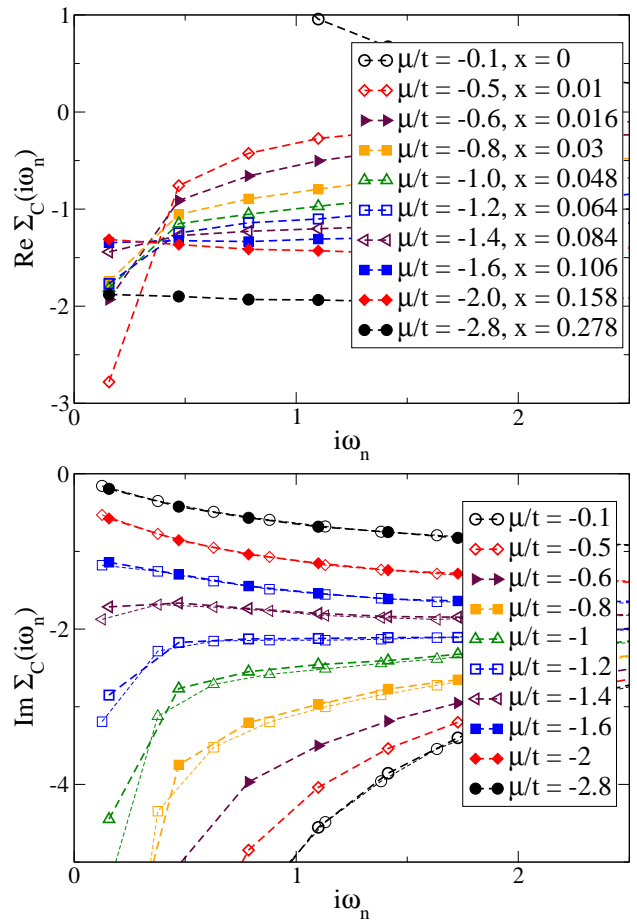

\begin{center}
\includegraphics[width=0.95\columnwidth]{ReSelfenergiesMuNegSectorC_tprime0.15.eps}\\
\includegraphics[width=0.95\columnwidth]{ImSelfenergiesMuNegSectorC_tprime0.15.eps}\\
\caption{Real (upper panel) and imaginary (lower panel) parts of sector $C$ self-energy calculated for indicated chemical potentials corresponding to hole doping, at $U/t = 7,$ $t'/t=-0.15,$ and $\beta t=20$ (thick lines, thick symbols) and $\beta t=25$ (thin lines, thin symbols). 
}\label{SigmaMuNegC.fig}
\end{center}
\end{figure}
\begin{figure}[t]
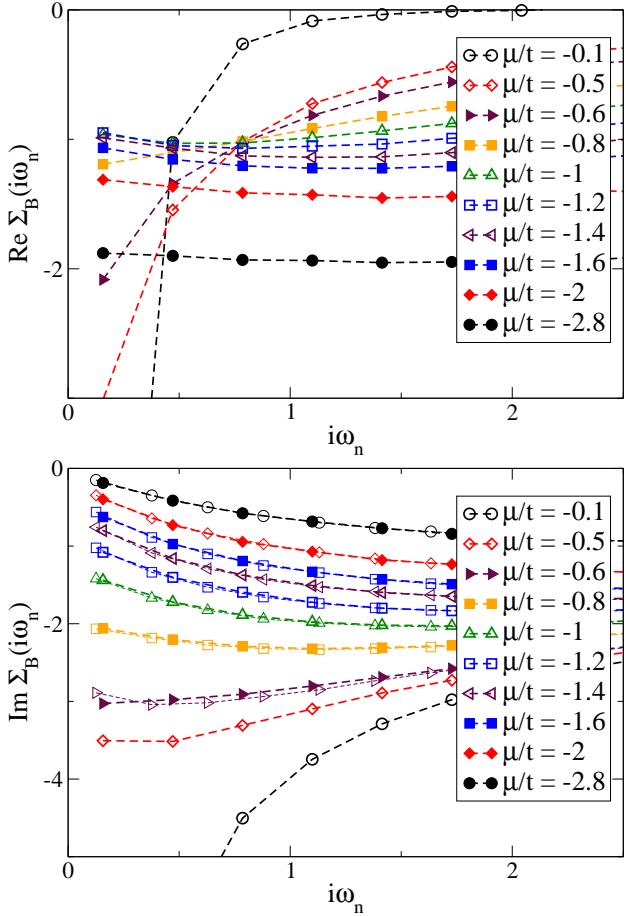

\begin{center}
\includegraphics[width=0.95\columnwidth]{ReSelfenergiesMuNegSectorB_tprime0.15.eps}
\includegraphics[width=0.95\columnwidth]{ImSelfenergiesMuNegSectorB_tprime0.15.eps}
\caption{Real (upper panel) and imaginary (lower panel) parts of sector $B$ self-energy calculated for indicated chemical potentials corresponding to hole doping, at $U/t = 7,$ $t'/t=-0.15,$ and $\beta t=20$ (thick lines, thick symbols) and $\beta t=25$ (thin lines, thin symbols).
}\label{SigmaMuNegB.fig}
\end{center}
\end{figure}

We now turn to the behavior of the self-energy. Figure~\ref{SigmaMuNegC.fig} shows the real and imaginary parts of the $C$-sector self-energies for the doping-driven transition.  As $\mu$ is increased above $\mu_C\approx -1.6t$ both real and imaginary parts develop the upturn expected if a pole appears in $\Sigma$. The temperature dependence given for $\text{Im}\Sigma$ confirms this interpretation. Figure~\ref{SigmaMuNegB.fig} shows the corresponding plots for sector $B$. While much lower temperatures (not accessible with presently available computer resources) would be required to establish the precise behavior, it is clear that for a range of $\mu$ around $-1$ $\text{Im}\Sigma_B$  does not seem to linearly extrapolate to zero while neither $\text{Im}\Sigma_B$ nor $\text{Re}\Sigma_B$ has a pole behavior. For $\mu \lesssim -1.5t$ Fermi-liquid behavior is apparently restored while for $\mu\gtrsim -0.6t$ a pole begins to develop.

\section{Spin correlations \label{Spins}}
\begin{figure}[tbh]
\begin{center}
\includegraphics[angle=0, width=0.95\columnwidth]{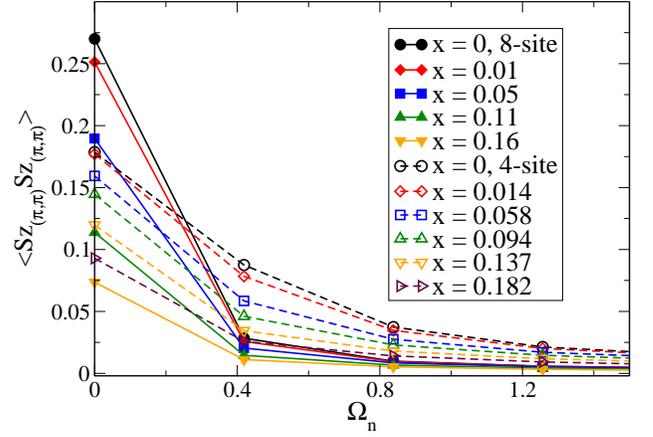}
\caption{$\langle Sz_K Sz_{K'}\rangle$ for $K=K'=(\pi,\pi)$ at $U/t = 7,$ $\beta t=15,$ and $t'/t = -0.15$ for the eight-site (bold lines, filled symbols) and four-site (dashed lines, empty symbols) cluster geometries, as a function of doping. 
}
\label{SzSz.fig}
\end{center}
\end{figure}
To obtain further insight into the physics of the sector-selective transition we present in Fig.~\ref{SzSz.fig} the cluster spin-correlation functions obtained from $S^z_{ij}(\tau-\tau')=\langle S^z_i(\tau) S^z_j(\tau')\rangle,$ where $i$ and $j$ are cluster site indices. While the cluster correlation functions are not identical to the spin correlators of the lattice model, they are expected to show the same physics. We have Fourier transformed the correlation functions into Matsubara frequency and cluster momentum-space. For all  cluster momentum sectors except $(\pi,\pi)$ the spin-spin correlation function is small (four times smaller than $S_{\pi,\pi}$ at $\mu=-2t$ and $20$ times smaller at $\mu=-0.5t$) and only weakly frequency and doping dependent. We therefore present in Fig.~\ref{SzSz.fig} only the $(\pi,\pi)$ correlators. For comparison we show also the same correlator for a four-site cluster. We see that as chemical potential is decreased through the sector-selective value $\mu_C \approx -1.6t$ the zero Matsubara frequency component grows substantially (by a factor of 4) and the first Matsubara frequency grows somewhat (factor of~$\approx 2$) while the other frequencies are essentially unchanged. Thus we see that the sector-selective regime is associated with a large value of quasiclassical $(\Omega_n=0)$ spin correlations. This finding is consistent with the arguments of Kyung {\it et al.}\cite{Kyung06} that the pseudogap is associated with backscattering arising from slow, reasonably long-ranged spin fluctuations.
By contrast, the correlation functions of the four-site cluster have a weaker frequency and doping dependence, as expected from the dominance of the singlet correlations on the plaquette.\cite{Gull08_plaquette}
However, it is not clear whether a spin fluctuations approach  can account for the pinning of the density to the commensurate value of $1/2$ that  we observe to be a hallmark of the sector-selective phenomenon in the eight-site cluster.

\section{Absence of Pomeranchuk effect\label{Pomeranchuk}}
\begin{figure}[tbh]
\begin{center}
\includegraphics[angle=0, width=0.95\columnwidth]{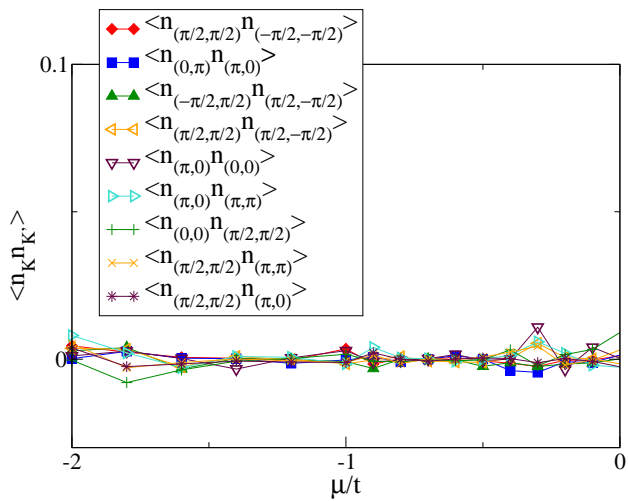}
\caption{Absence of Pomeranchuk effect: density-density correlators for $U/t = 7,$ $\beta t=15,$ and  $t'/t=-0.3$ for momenta $K \neq K'.$ All correlators are
zero within error bars.
}
\label{pomeranchuk.fig}
\end{center}
\end{figure}

%\subsection{Absence of Pomeranchuk effect}\label{Physics}
A nematic instability\cite{Pomeranchuk59} corresponding to a breaking of rotation symmetry without a corresponding long-ranged order has been  extensively discussed in the context of strongly correlated systems.\cite{Oganesyan01,Kivelson03,Kee04,Kee05,Borzi07,Fradkin07,Wu07,DellAnna07} For example, in the context of the metamagnetic material Sr$_3$Ru$_2$O$_7$ a transition corresponding to an increase in density of electrons with momentum near $(\pi,0)$ and corresponding decrease in the density of electrons with momentum near $(0, \pi)$ (or the reverse) has been discussed.\cite{Kee04} The locking of the sector density to the commensurate value $n_C=1/2$ in the sector-selective region motivates the question whether the nematic fluctuations are particularly large near the sector-selective phase. A tendency toward nematicity would correspond to an anticorrelation of density fluctuations between different symmetry equivalent sectors. Figure~\ref{pomeranchuk.fig} shows the results of our measurements of the appropriate correlation function as a function of chemical potential. 
We see (within errors) no correlations, even near the sector-selective point. We therefore conclude that within the eight-site DCA approximation the Hubbard model does not exhibit any tendency toward charge nematic behavior at the intermediate couplings we have studied. This  finding is in disagreement with RG studies of the Hubbard model \cite{Halboth00,Hankevych02} and simulations of the $t$-$J$ model,\cite{Yamase00,Yamase00B,Edegger06} where a Pomeranchuk instability is seen. Whether the discrepancies are a consequence of  a difference in parameter regimes examined or arise from differences in approximation methods remains  an open question.
%We see (within errors) no correlations, even near the sector-selective point. We therefore conclude that the model does not exhibit any tendency towards charge nematic behavior.

\section{Relation to Orbital Selective Mott Transition \label{Selective}}

In the DCA implementation of cluster dynamical mean-field theory, the impurity-model Green's function is diagonalized in a basis labeled by momentum quantum numbers $K$ which are related, via a self-consistency condition, to averages over particular regions of the Brillouin-zone of the lattice model.  The model may be viewed more abstractly, regarding the labels $K$ as orbitals of a general multi-level impurity model and averages over different regions of the Brillouin-zone as arising from different baths. In this way, as noted by Biermann {\it et al.},\cite{Biermann05b} the DCA equations may be mapped onto the general problem of a multilevel impurity model. The sector-selective transitions studied here are then simply different instantiations of the general phenomenon of the ``orbital-selective Mott transition,'' extensively discussed in the context of transition-metal oxides \cite{Anisimov02,Koga04,Biermann05b,Koga05,Werner072level,Werner093level,Jakobi09} 
with partially filled $d$ levels. This connection was explored by Ferrero and co-workers \cite{Ferrero09,Ferrero09b} and  also by Liebsch \cite{Liebsch08} and provides both insights into the physics of the cluster dynamical mean-field theory and insights back to the single-site physics of multiorbital transition-metal oxides. 

However, we note that when expressed in the orbital basis the interactions in the cluster studied here are considerably more complicated than those normally studied in the orbitally degenerate transition-metal oxide case.  Liebsch {\it et al.} found that the four-site clusters they studied did not exhibit momentum-selective transitions. The presence of two interaction-driven transitions at half filling seems (as noted above) to follow naturally from the different bandwidths of the two relevant sectors while the persistence of the orbital-selective effect on doping is similar to the two-orbital results presented in Ref.~\onlinecite{Werner072level}. We note, however, that  neither  the second-order nature of the first transition nor   the collapse of the two transitions into one as the $t'$ parameter is increased seem to have been noticed in the orbital-selective Mott-transition literature.  These issues seem worth exploring in the latter context.  Further, the essential role apparently played in our calculations by the pinning of the density to $1/2$ in the insulating orbitals requires further investigation. In the two-site calculations of Ferrero {\it et al.} this pinning was not found.

From the point of view of the orbital-selective Mott transition the models studied here have a special feature, which is that the different ``orbital'' ($K$) sectors are decoupled both by the symmetry of the impurity model and by the self-consistency condition. In other cluster DMFT implementations (for example, cellular dynamical mean-field theory) this may not be the case. In the context of the orbital-selective Mott phenomenon  this decoupling has been shown to have an important consequence; it rules out a Kondo coupling between the gapped and the ungapped orbital which, if antiferromagnetic, might destabilize the orbital-selective phase at temperatures below the Kondo scale.\cite{Biermann05b} It is therefore natural to ask if the orbital-selective phase we find survives beyond the DCA approximation we have used here. We note that the transitions we find in the cluster DMFT context have an additional feature not yet explored in the orbital-selective literature the orbital which first becomes insulating (sector $C$) is two fold degenerate. The insulating orbitals may therefore have a more complex spin structure than in the non-degenerate case, and, in particular, it seems likely that the singlet state of the two orbitals will be favored, in which case a Kondo effect would not necessarily occur. Cellular dynamical mean-field theory calculations (in which the self-consistency condition couples the sectors)  would be very valuable, and it may be interesting to explore the sector degeneracy effects further in the selective Mott-transition literature.

\section{Summary \label{Summary}}
Building on previous results \cite{Werner098site} we have shown  in this paper that the  eight-site DCA approximation to the intermediate-coupling Hubbard model yields results which exhibit a remarkable qualitative correspondence to the pseudogap effects occurring in high-temperature superconductors. For parameters relevant to high-$T_c$ superconductors we find that hole doping
but not electron doping leads to a two-stage metal-insulator transition in which a gap opens first around the $(\pi,0)$ point. Our conclusions are based on an analysis of directly measured quantities  and do not involve analytical continuation or interpolation of self-energies. In this concluding section we summarize our results and place them in the context of other studies of high-temperature superconductivity and of other dynamical mean-field analyses. This section includes pointers back to relevant sections of the text and may be  read independently of the rest of the paper. 

A crucial finding of Ref.~\onlinecite{Werner098site} was that for  coupling strong enough that the model is insulating at $n=1$ a two-stage metal-insulator transition occurs as  the insulating phase is approached from the regime of large hole doping. As the doping is reduced below  a critical doping $x_c$ ($\sim 0.10$ for $U=7t$  and increasing weakly as $U$ is increased)
a gap appears in the momentum sector containing the ``antinodal'' region $(0,\pi)$ while the states in the nodal region remain gapless until the half-filled point is reached.
The resolution available to us does not permit us to distinguish between a full gap (density of states $=0$ for a nonzero range of energies) and a soft or pseudogap where the density of states vanishes only at one frequency or is strongly suppressed over a range of frequencies but does not vanish.
We found that in this ``sector-selective'' region the antinodal gap is small in comparison to the correlation gap found at half filling, although it does grow rapidly  as doping is decreased. The small gap magnitude but rapid increase can be seen from the relatively weak temperature dependence %\textcolor{red}{PW: WHY SHOULD A WEAK T DEPENDENCE INDICATE A SMALL GAP? ISN'T IT JUST THE OPPOSITE} 
of the many-body density of states (Fig.~\ref{ghalf_mu_tprime0.15.fig}) and from the Matsubara-axis self-energies (Figs.~\ref{SigmaMuNegB.fig} and~\ref{SigmaMuNegC.fig}). For an electronic dispersion with particle-hole symmetry the two-stage transition occurs for both electron and hole doping but for particle-hole asymmetries of the magnitude  believed \cite{Andersen95} to be relevant for high-temperature superconductors we find that the two-stage transition occurs only on the hole-doping side (Fig.~\ref{PDDoping}). These two findings are the essence of the high-$T_c$ pseudogap  phenomenon and thus our results add support to the widely held belief that the intermediate-coupling Hubbard model contains much if not all of the  essential physics of high-$T_c$ superconductivity. 

The ``antinodal gap'' phase appears to be non-Fermi-liquid in two senses; of course the gapping of the antinodal sector implies that the Fermi surface 
violates the Luttinger theorem associated with the full carrier density. Further, ungapped states in the nodal sector are non-Fermi-liquid in 
the sense that the many-body density of states appears  (Fig.~\ref{ghalf_mu_tprime0.15.fig}) to assume a low-temperature value which is neither 
the non-interacting value (as expected if the states in this sector were Fermi-liquid like) nor vanishing (as expected if the sector were gapped) 
and in the sense that the imaginary part of the Matsubara-axis self-energy appears not to vanish as $i\omega_n\rightarrow 0$ at the temperatures 
accessible to us (Fig.~\ref{SigmaMuNegB.fig}).  

The antinodal gap regime is separated  from a large-doping Fermi-liquid phase  by a zero-temperature transition whose existence is inferred from a crossing of the temperature dependence of the many-body density of states. The transition is apparently continuous, characterized by a gap which opens smoothly as the doping is decreased below the apparent quantum critical point (Figs.~\ref{inv_interact_0.fig} and \ref{SigmaMuNegC.fig}). Because the transition does not involve an order parameter it is a smooth crossover at $T>0$.  An apparently rather similar transition occurs if the interaction is varied at density $n=1$ and the particle-hole symmetry breaking is not too large (Figs.~\ref{ghalf_interaction.fig} and~\ref{inv_interact_0.fig}).

The second transition, at which the nodal sector develops a gap, is first order (characterized by discontinuities in the gap amplitude) in the interaction-driven case. In the doping-driven case the behavior of physical quantities appears smooth but the chemical potential is tuned into a large-amplitude pre-existing gap in the nodal sector (see Fig.~\ref{SigmaMuNegB.fig}). 

An antinodal gap which varies smoothly with doping and vanishes above a critical doping level, is an essential feature of the pseudogap observed in hole-doped high-$T_c$ materials \cite{Hufner08}.
Interestingly, the many-body density of states obtained by analytical continuation of quantum Monte Carlo data for four-site clusters (which do not provide the momentum resolution needed to distinguish the nodal and antinodal sectors) revealed in the doped case a small gap  pinned to the Fermi level.\cite{Gull08_plaquette} Ferrero {\it et al.}\cite{Ferrero09,Ferrero09b}  subsequently found a similar feature in a two-site calculation. We speculate that this is the expression, in the smaller clusters, of the two-stage transition which is clearly revealed in the larger eight-site cluster. Further investigation of the relation between two-, four-, and eight-site cluster results is in progress. 

The transition at which sector $C$ becomes gapped appears to be different from the quantum phase transition identified by Vidhyadhiraja {\it et al.}\cite{Vidhyadhiraja09} and the related phenomena studied by Liebsch \cite{Liebsch09} using a four-site cluster. The Vidhyadhiraja transition  occurs at the higher doping $x\sim 0.15$ (corresponding to chemical potential $\mu\approx -2t$) where within our resolution no effects occur. However, we note that the transition identified by Vidhyadhiraja {\it et al.} was obtained for a different cluster geometry and more importantly involves a change in the self-energy from a Fermi-liquid to a marginal Fermi-liquid form; such a change  would lead only to subtle effects in the quantities we are able to measure accurately. The range of temperatures we can access may also be too small.

It is important to note that the transition is not driven by the van-Hove singularity in the density of states. As can be seen from Fig.~\ref{mueff.fig}, at $t'=-0.15t$ the transition to the antinodal gap phase occurs {\em after} the renormalized Fermi level has passed through the Van Hove point.  A generic feature of weak-coupling theories based on perturbation about a Fermi-liquid phase is the importance of Van Hove physics. The absence of a clear association with the Van Hove point identifies the transition we find as an intermediate or strong-coupling phenomenon. 

The sector-selective transitions that we find are associated with a pinning of the sector occupancy to the half-filled value $n_\text{sector}=1/2$. The pinning is a $T\rightarrow 0$ effect (consistent with the lack of an order parameter and nonexistence of a $T>0$ transition) and because of the small gaps the approach to the $T=0$ limit can be slow.  The pinning of the density and the slow approach to the $T=0$ value can be seen in  Fig.~\ref {densities_mu_tprime0.15.fig}. The importance of the pinning of the density is also revealed by the dependence of our findings on the particle-hole symmetry breaking parameter $t'$.  The sector-selective phenomenon occurs both as the interaction strength is varied at density $n=1$ (Fig.~\ref{PDHalffilled}) and as the density is varied at strong interaction (Fig.~\ref{PDDoping}). 

In the interaction-driven case a two-stage transition occurs only for $|t'|$ sufficiently small. At $t'=0$ and $n=1$ both the nodal and antinodal sectors are half filled at all values of the interaction strength and the smaller bandwidth of the antinodal sector (Fig.~\ref{nonint_diff.fig}) explains why this sector becomes insulating at a smaller interaction strength than does the nodal sector.  However, for $t'\neq 0$ in the small $U$ fully metallic phase both sectors have occupancy different from the half-filled value while in the sector-selective phase the gapped sector (and therefore also the ungapped one)  have density equal to the half-filled value. As $U$ is increased , momentum differentiation (a difference in electron self-energy between sectors) causes charge to be transferred between sectors. Thus the sector-selective phase is energetically disfavored by the need to overcome the kinetic energy and equalize the charge distribution between sectors. At small $|t'|$ this effect leads only to a weak increase in the critical interactions strength but as $|t'|$ is increased the critical $U$ required to equalize the charge distribution becomes larger than the value needed to open a gap in the nodal sector, leading to one first-order transition.

A very similar phenomenon may be seen in the doping-driven transition (Fig.~\ref{PDDoping}). For small $|t'|$ sector-selective transitions occur for both electron and hole-doping but as $t'/t$ becomes more negative (positive), the transitions on the electron-doped (hole-doped) side merge and become first order, again because the density difference between sectors becomes too large to sustain without gapping both sectors. This physics is revealed most clearly in Fig.~\ref{densities_mu_tprime0.15.fig}, where for  $t'/t = -0.3$ we  see that as the chemical potential is raised across the insulating gap the density of the $B$ sector actually drops below the half-filling value even though the mean electron density increases.  Similarly, on the hole-doped side the sector-selective transition occurs when the chemical potential reaches the value at which the antinodal sector becomes half filled.

The close association between commensurate filling and the antinodal gap leads us to identify the antinodal transition as being of the Mott type. We also observe in passing that no evidence for ``nematic'' density fluctuations or a Pomeranchuk-type Fermi surface instability is found (Fig.~\ref{pomeranchuk.fig}).  If the transition were driven exclusively by short-ranged antiferromagnetic order there would, as far as we are aware, be no reason for the sector density to be pinned. We observe, however, that the onset of the sector-selective phase is associated with a strong increase in quasi thermal, long correlation length antiferromagnetic spin fluctuations (Fig.~\ref{SzSz.fig}). As in the single-site dynamical mean-field theory, the onset of a Mott phase is associated with local-moment formation and the presence of the local moments will of course lead to strongly enhanced spin correlations. Interestingly, we see no obvious sign of a spin gap. While one expects the eight-site cluster to have a low-spin ground state, the energy difference between lowest-lying singlet and triplet states is evidently below our temperature resolution. 

In summary, we have presented evidence that the intermediate-coupling Hubbard model exhibits a momentum-selective, multistage Mott transition leading to a pseudogap phenomenon in remarkable qualitative agreement with high-$T_c$ materials. The systematics of our findings strongly suggest that a weak-coupling interpretation is not tenable.  Important open questions, presently under investigation, include the relation of the present eight-site cluster results to those obtained on smaller and on larger clusters. In this context we note that studies of  non particle-hole symmetric clusters (either doped or with $t'\neq0$) encounter a fermion sign problem which becomes progressively more severe for lower temperatures, stronger interactions, and larger clusters so unless some remarkable new methodological innovation appears extension of our results too much beyond the parameters studied here will not be possible.  However, working with presently accessible cluster sizes and interaction strengths, an extension of our results to include long-ranged magnetic and superconducting order is possible and is an important open question. 

Recently, we became aware of work of Khatami et al.,\cite{Khatami09} which reported DCA calculations of the compressibility $dn/d\mu$ indicating a sharp first-order transition with associated hysteresis at a small electron doping for $|t'|\gtrsim 0.2$. We associate this transition with our findings (Fig.~\ref{densities_mu_tprime0.15.fig} and \ref{PDDoping}) that for $|t'|  \gtrsim 0.2$ and large $U$ the doping-driven metal-insulator transition is first order on the electron doped side. We expect that this first-order transition will lead to hysteresis and anomalous behavior of $dn/d\mu$ at low dopings, on the order of the jump in density shown in Fig.~\ref{ntotvsmu.fig}.

\acknowledgments{Calculations have been performed on the Brutus cluster at ETH Zurich and using HPC resources from GENCI-CCRT (Grant No. 2009-t2009056112). E.~G. and A.~J.~M. are supported by NSF under Grant No.~DMR-0705847, P.~W. by the Swiss National Science Foundation  (Grant No.~PP002-118866). The codes used are based on the ALPS (Ref.~\onlinecite{ALPS}) library. We thank A.~Georges and M.~Ferrero for helpful conversations.}
\bibliography{refs_small}

\begin{thebibliography}{54}
\expandafter\ifx\csname natexlab\endcsname\relax\def\natexlab#1{#1}\fi
\expandafter\ifx\csname bibnamefont\endcsname\relax
  \def\bibnamefont#1{#1}\fi
\expandafter\ifx\csname bibfnamefont\endcsname\relax
  \def\bibfnamefont#1{#1}\fi
\expandafter\ifx\csname citenamefont\endcsname\relax
  \def\citenamefont#1{#1}\fi
\expandafter\ifx\csname url\endcsname\relax
  \def\url#1{\texttt{#1}}\fi
\expandafter\ifx\csname urlprefix\endcsname\relax\def\urlprefix{URL }\fi
\providecommand{\bibinfo}[2]{#2}
\providecommand{\eprint}[2][]{\url{#2}}

\bibitem[{\citenamefont{Maier et~al.}(2005)\citenamefont{Maier, Jarrell,
  Pruschke, and Hettler}}]{Maier05}
\bibinfo{author}{\bibfnamefont{T.}~\bibnamefont{Maier}},
  \bibinfo{author}{\bibfnamefont{M.}~\bibnamefont{Jarrell}},
  \bibinfo{author}{\bibfnamefont{T.}~\bibnamefont{Pruschke}}, \bibnamefont{and}
  \bibinfo{author}{\bibfnamefont{M.~H.} \bibnamefont{Hettler}},
  \bibinfo{journal}{Rev. Mod. Phys.} \textbf{\bibinfo{volume}{77}},
  \bibinfo{pages}{1027} (\bibinfo{year}{2005}).

\bibitem[{\citenamefont{Jarrell et~al.}(2001)\citenamefont{Jarrell, Maier,
  Hettler, and Tahvildarzadeh}}]{Jarrell01}
\bibinfo{author}{\bibfnamefont{M.}~\bibnamefont{Jarrell}},
  \bibinfo{author}{\bibfnamefont{T.}~\bibnamefont{Maier}},
  \bibinfo{author}{\bibfnamefont{M.~H.} \bibnamefont{Hettler}},
  \bibnamefont{and} \bibinfo{author}{\bibfnamefont{A.~N.}
  \bibnamefont{Tahvildarzadeh}}, \bibinfo{journal}{Europhys. Lett.}
  \textbf{\bibinfo{volume}{56}}, \bibinfo{pages}{563} (\bibinfo{year}{2001}).

\bibitem[{\citenamefont{Parcollet et~al.}(2004)\citenamefont{Parcollet, Biroli,
  and Kotliar}}]{Parcollet04}
\bibinfo{author}{\bibfnamefont{O.}~\bibnamefont{Parcollet}},
  \bibinfo{author}{\bibfnamefont{G.}~\bibnamefont{Biroli}}, \bibnamefont{and}
  \bibinfo{author}{\bibfnamefont{G.}~\bibnamefont{Kotliar}},
  \bibinfo{journal}{Phys. Rev. Lett.} \textbf{\bibinfo{volume}{92}},
  \bibinfo{pages}{226402} (\bibinfo{year}{2004}).

\bibitem[{\citenamefont{Civelli et~al.}(2005)\citenamefont{Civelli, Capone,
  Kancharla, Parcollet, and Kotliar}}]{Civelli05}
\bibinfo{author}{\bibfnamefont{M.}~\bibnamefont{Civelli}},
  \bibinfo{author}{\bibfnamefont{M.}~\bibnamefont{Capone}},
  \bibinfo{author}{\bibfnamefont{S.~S.} \bibnamefont{Kancharla}},
  \bibinfo{author}{\bibfnamefont{O.}~\bibnamefont{Parcollet}},
  \bibnamefont{and} \bibinfo{author}{\bibfnamefont{G.}~\bibnamefont{Kotliar}},
  \bibinfo{journal}{Phys. Rev. Lett.} \textbf{\bibinfo{volume}{95}},
  \bibinfo{pages}{106402} (\bibinfo{year}{2005}).

\bibitem[{\citenamefont{Kyung et~al.}(2006)\citenamefont{Kyung, Kancharla,
  S\'{e}n\'{e}chal et~al.}}]{Kyung06}
\bibinfo{author}{\bibfnamefont{B.}~\bibnamefont{Kyung}},
  \bibinfo{author}{\bibfnamefont{S.~S.} \bibnamefont{Kancharla}},
  \bibinfo{author}{\bibfnamefont{D.}~\bibnamefont{S\'{e}n\'{e}chal}},
  \bibnamefont{et~al.}, \bibinfo{journal}{Phys. Rev. B}
  \textbf{\bibinfo{volume}{73}}, \bibinfo{eid}{165114} (\bibinfo{year}{2006}).

\bibitem[{\citenamefont{Macridin
  et~al.}(2006{\natexlab{a}})\citenamefont{Macridin, Jarrell, Maier, Kent, and
  D'Azevedo}}]{Macridin06}
\bibinfo{author}{\bibfnamefont{A.}~\bibnamefont{Macridin}},
  \bibinfo{author}{\bibfnamefont{M.}~\bibnamefont{Jarrell}},
  \bibinfo{author}{\bibfnamefont{T.}~\bibnamefont{Maier}},
  \bibinfo{author}{\bibfnamefont{P.~R.~C.} \bibnamefont{Kent}},
  \bibnamefont{and}
  \bibinfo{author}{\bibfnamefont{E.}~\bibnamefont{D'Azevedo}},
  \bibinfo{journal}{Phys. Rev. Lett.} \textbf{\bibinfo{volume}{97}},
  \bibinfo{pages}{036401} (\bibinfo{year}{2006}{\natexlab{a}}).

\bibitem[{\citenamefont{Chakraborty et~al.}(2008)\citenamefont{Chakraborty,
  Galanakis, and Phillips}}]{Chakraborty08}
\bibinfo{author}{\bibfnamefont{S.}~\bibnamefont{Chakraborty}},
  \bibinfo{author}{\bibfnamefont{D.}~\bibnamefont{Galanakis}},
  \bibnamefont{and} \bibinfo{author}{\bibfnamefont{P.}~\bibnamefont{Phillips}},
  \bibinfo{journal}{Phys. Rev. B} \textbf{\bibinfo{volume}{78}},
  \bibinfo{pages}{212504} (\bibinfo{year}{2008}).

\bibitem[{\citenamefont{Haule}(2007)}]{Haule07}
\bibinfo{author}{\bibfnamefont{K.}~\bibnamefont{Haule}},
  \bibinfo{journal}{Phys. Rev. B} \textbf{\bibinfo{volume}{75}},
  \bibinfo{pages}{155113} (\bibinfo{year}{2007}).

\bibitem[{\citenamefont{Park et~al.}(2008)\citenamefont{Park, Haule, and
  Kotliar}}]{Park08plaquette}
\bibinfo{author}{\bibfnamefont{H.}~\bibnamefont{Park}},
  \bibinfo{author}{\bibfnamefont{K.}~\bibnamefont{Haule}}, \bibnamefont{and}
  \bibinfo{author}{\bibfnamefont{G.}~\bibnamefont{Kotliar}},
  \bibinfo{journal}{Phys. Rev. Lett.} \textbf{\bibinfo{volume}{101}},
  \bibinfo{pages}{186403} (\bibinfo{year}{2008}).

\bibitem[{\citenamefont{Gull et~al.}(2008{\natexlab{a}})\citenamefont{Gull,
  Werner, Wang, Troyer, and Millis}}]{Gull08_plaquette}
\bibinfo{author}{\bibfnamefont{E.}~\bibnamefont{Gull}},
  \bibinfo{author}{\bibfnamefont{P.}~\bibnamefont{Werner}},
  \bibinfo{author}{\bibfnamefont{X.}~\bibnamefont{Wang}},
  \bibinfo{author}{\bibfnamefont{M.}~\bibnamefont{Troyer}}, \bibnamefont{and}
  \bibinfo{author}{\bibfnamefont{A.~J.} \bibnamefont{Millis}},
  \bibinfo{journal}{Europhys. Lett.)} \textbf{\bibinfo{volume}{84}},
  \bibinfo{pages}{37009} (\bibinfo{year}{2008}{\natexlab{a}}).

\bibitem[{\citenamefont{Stanescu and Kotliar}(2006)}]{Stanescu06}
\bibinfo{author}{\bibfnamefont{T.~D.} \bibnamefont{Stanescu}} \bibnamefont{and}
  \bibinfo{author}{\bibfnamefont{G.}~\bibnamefont{Kotliar}},
  \bibinfo{journal}{Phys. Rev. B} \textbf{\bibinfo{volume}{74}},
  \bibinfo{pages}{125110} (\bibinfo{year}{2006}).

\bibitem[{\citenamefont{Ferrero
  et~al.}(2009{\natexlab{a}})\citenamefont{Ferrero, Cornaglia, {De~Leo},
  Parcollet, Kotliar, and Georges}}]{Ferrero09}
\bibinfo{author}{\bibfnamefont{M.}~\bibnamefont{Ferrero}},
  \bibinfo{author}{\bibfnamefont{P.~S.} \bibnamefont{Cornaglia}},
  \bibinfo{author}{\bibfnamefont{L.}~\bibnamefont{{De~Leo}}},
  \bibinfo{author}{\bibfnamefont{O.}~\bibnamefont{Parcollet}},
  \bibinfo{author}{\bibfnamefont{G.}~\bibnamefont{Kotliar}}, \bibnamefont{and}
  \bibinfo{author}{\bibfnamefont{A.}~\bibnamefont{Georges}},
  \bibinfo{journal}{Europhys. Lett.)} \textbf{\bibinfo{volume}{85}},
  \bibinfo{pages}{57009} (\bibinfo{year}{2009}{\natexlab{a}}).

\bibitem[{\citenamefont{Ferrero
  et~al.}(2009{\natexlab{b}})\citenamefont{Ferrero, Cornaglia, {De~Leo},
  Parcollet, Kotliar, and Georges}}]{Ferrero09b}
\bibinfo{author}{\bibfnamefont{M.}~\bibnamefont{Ferrero}},
  \bibinfo{author}{\bibfnamefont{P.~S.} \bibnamefont{Cornaglia}},
  \bibinfo{author}{\bibfnamefont{L.}~\bibnamefont{{De~Leo}}},
  \bibinfo{author}{\bibfnamefont{O.}~\bibnamefont{Parcollet}},
  \bibinfo{author}{\bibfnamefont{G.}~\bibnamefont{Kotliar}}, \bibnamefont{and}
  \bibinfo{author}{\bibfnamefont{A.}~\bibnamefont{Georges}},
  \bibinfo{journal}{Phys. Rev. B} \textbf{\bibinfo{volume}{80}},
  \bibinfo{pages}{064501} (\bibinfo{year}{2009}{\natexlab{b}}).

\bibitem[{\citenamefont{Sakai et~al.}(2009)\citenamefont{Sakai, Motome, and
  Imada}}]{Sakai09}
\bibinfo{author}{\bibfnamefont{S.}~\bibnamefont{Sakai}},
  \bibinfo{author}{\bibfnamefont{Y.}~\bibnamefont{Motome}}, \bibnamefont{and}
  \bibinfo{author}{\bibfnamefont{M.}~\bibnamefont{Imada}},
  \bibinfo{journal}{Phys. Rev. Lett.} \textbf{\bibinfo{volume}{102}},
  \bibinfo{pages}{056404} (\bibinfo{year}{2009}).

\bibitem[{\citenamefont{Liebsch and Tong}(2009)}]{Liebsch09}
\bibinfo{author}{\bibfnamefont{A.}~\bibnamefont{Liebsch}} \bibnamefont{and}
  \bibinfo{author}{\bibfnamefont{N.-H.} \bibnamefont{Tong}},
  \bibinfo{journal}{Phys. Rev. B} \textbf{\bibinfo{volume}{80}},
  \bibinfo{pages}{165126} (\bibinfo{year}{2009}).

\bibitem[{\citenamefont{Haule and Kotliar}(2007)}]{Haule07c}
\bibinfo{author}{\bibfnamefont{K.}~\bibnamefont{Haule}} \bibnamefont{and}
  \bibinfo{author}{\bibfnamefont{G.}~\bibnamefont{Kotliar}},
  \bibinfo{journal}{Phys. Rev. B} \textbf{\bibinfo{volume}{76}},
  \bibinfo{pages}{092503} (\bibinfo{year}{2007}).

\bibitem[{\citenamefont{Vidhyadhiraja et~al.}(2009)\citenamefont{Vidhyadhiraja,
  Macridin, \c{S}en, Jarrell, and Ma}}]{Vidhyadhiraja09}
\bibinfo{author}{\bibfnamefont{N.~S.} \bibnamefont{Vidhyadhiraja}},
  \bibinfo{author}{\bibfnamefont{A.}~\bibnamefont{Macridin}},
  \bibinfo{author}{\bibfnamefont{C.}~\bibnamefont{\c{S}en}},
  \bibinfo{author}{\bibfnamefont{M.}~\bibnamefont{Jarrell}}, \bibnamefont{and}
  \bibinfo{author}{\bibfnamefont{M.}~\bibnamefont{Ma}}, \bibinfo{journal}{Phys.
  Rev. Lett.} \textbf{\bibinfo{volume}{102}}, \bibinfo{pages}{206407}
  (\bibinfo{year}{2009}).

\bibitem[{\citenamefont{Werner et~al.}(2009{\natexlab{a}})\citenamefont{Werner,
  Gull, Parcollet, and Millis}}]{Werner098site}
\bibinfo{author}{\bibfnamefont{P.}~\bibnamefont{Werner}},
  \bibinfo{author}{\bibfnamefont{E.}~\bibnamefont{Gull}},
  \bibinfo{author}{\bibfnamefont{O.}~\bibnamefont{Parcollet}},
  \bibnamefont{and} \bibinfo{author}{\bibfnamefont{A.~J.}
  \bibnamefont{Millis}}, \bibinfo{journal}{Phys. Rev. B}
  \textbf{\bibinfo{volume}{80}}, \bibinfo{pages}{045120}
  (\bibinfo{year}{2009}{\natexlab{a}}).

\bibitem[{\citenamefont{Biermann et~al.}(2005)\citenamefont{Biermann, de'
  Medici, and Georges}}]{Biermann05b}
\bibinfo{author}{\bibfnamefont{S.}~\bibnamefont{Biermann}},
  \bibinfo{author}{\bibfnamefont{L.}~\bibnamefont{de' Medici}},
  \bibnamefont{and} \bibinfo{author}{\bibfnamefont{A.}~\bibnamefont{Georges}},
  \bibinfo{journal}{Phys. Rev. Lett.} \textbf{\bibinfo{volume}{95}},
  \bibinfo{pages}{206401} (\bibinfo{year}{2005}).

\bibitem[{\citenamefont{Anisimov et~al.}(2002)\citenamefont{Anisimov, Nekrasov,
  Kondakov, Rice, and Sigrist}}]{Anisimov02}
\bibinfo{author}{\bibfnamefont{V.}~\bibnamefont{Anisimov}},
  \bibinfo{author}{\bibfnamefont{I.}~\bibnamefont{Nekrasov}},
  \bibinfo{author}{\bibfnamefont{D.}~\bibnamefont{Kondakov}},
  \bibinfo{author}{\bibfnamefont{T.}~\bibnamefont{Rice}}, \bibnamefont{and}
  \bibinfo{author}{\bibfnamefont{M.}~\bibnamefont{Sigrist}},
  \bibinfo{journal}{Eur. Phys. J. B} \textbf{\bibinfo{volume}{25}},
  \bibinfo{pages}{191} (\bibinfo{year}{2002}).

\bibitem[{\citenamefont{Koga et~al.}(2004)\citenamefont{Koga, Kawakami, Rice,
  and Sigrist}}]{Koga04}
\bibinfo{author}{\bibfnamefont{A.}~\bibnamefont{Koga}},
  \bibinfo{author}{\bibfnamefont{N.}~\bibnamefont{Kawakami}},
  \bibinfo{author}{\bibfnamefont{T.~M.} \bibnamefont{Rice}}, \bibnamefont{and}
  \bibinfo{author}{\bibfnamefont{M.}~\bibnamefont{Sigrist}},
  \bibinfo{journal}{Phys. Rev. Lett.} \textbf{\bibinfo{volume}{92}},
  \bibinfo{pages}{216402} (\bibinfo{year}{2004}).

\bibitem[{\citenamefont{Liebsch}(2005)}]{Liebsch05}
\bibinfo{author}{\bibfnamefont{A.}~\bibnamefont{Liebsch}},
  \bibinfo{journal}{Phys. Rev. Lett.} \textbf{\bibinfo{volume}{95}},
  \bibinfo{eid}{116402} (\bibinfo{year}{2005}).

\bibitem[{\citenamefont{P\'{e}pin}(2008)}]{Pepin08}
\bibinfo{author}{\bibfnamefont{C.}~\bibnamefont{P\'{e}pin}},
  \bibinfo{journal}{Phys. Rev. B} \textbf{\bibinfo{volume}{77}},
  \bibinfo{eid}{245129} (\bibinfo{year}{2008}).

\bibitem[{\citenamefont{Liebsch et~al.}(2008)\citenamefont{Liebsch, Ishida, and
  Merino}}]{Liebsch08}
\bibinfo{author}{\bibfnamefont{A.}~\bibnamefont{Liebsch}},
  \bibinfo{author}{\bibfnamefont{H.}~\bibnamefont{Ishida}}, \bibnamefont{and}
  \bibinfo{author}{\bibfnamefont{J.}~\bibnamefont{Merino}},
  \bibinfo{journal}{Phys. Rev. B} \textbf{\bibinfo{volume}{78}},
  \bibinfo{eid}{165123} (\bibinfo{year}{2008}).

\bibitem[{\citenamefont{Macridin and Jarrell}(2008)}]{Macridin08}
\bibinfo{author}{\bibfnamefont{A.}~\bibnamefont{Macridin}} \bibnamefont{and}
  \bibinfo{author}{\bibfnamefont{M.}~\bibnamefont{Jarrell}},
  \bibinfo{journal}{Phys. Rev. B} \textbf{\bibinfo{volume}{78}},
  \bibinfo{eid}{241101} (\bibinfo{year}{2008}).

\bibitem[{\citenamefont{Hettler et~al.}(1998)\citenamefont{Hettler,
  Tahvildar-Zadeh, Jarrell, Pruschke, and Krishnamurthy}}]{Hettler98}
\bibinfo{author}{\bibfnamefont{M.~H.} \bibnamefont{Hettler}},
  \bibinfo{author}{\bibfnamefont{A.~N.} \bibnamefont{Tahvildar-Zadeh}},
  \bibinfo{author}{\bibfnamefont{M.}~\bibnamefont{Jarrell}},
  \bibinfo{author}{\bibfnamefont{T.}~\bibnamefont{Pruschke}}, \bibnamefont{and}
  \bibinfo{author}{\bibfnamefont{H.~R.} \bibnamefont{Krishnamurthy}},
  \bibinfo{journal}{Phys. Rev. B} \textbf{\bibinfo{volume}{58}},
  \bibinfo{pages}{R7475} (\bibinfo{year}{1998}).

\bibitem[{\citenamefont{Georges et~al.}(1996)\citenamefont{Georges, Kotliar,
  Krauth, and Rozenberg}}]{Georges96}
\bibinfo{author}{\bibfnamefont{A.}~\bibnamefont{Georges}},
  \bibinfo{author}{\bibfnamefont{G.}~\bibnamefont{Kotliar}},
  \bibinfo{author}{\bibfnamefont{W.}~\bibnamefont{Krauth}}, \bibnamefont{and}
  \bibinfo{author}{\bibfnamefont{M.~J.} \bibnamefont{Rozenberg}},
  \bibinfo{journal}{Rev. Mod. Phys.} \textbf{\bibinfo{volume}{68}},
  \bibinfo{pages}{13} (\bibinfo{year}{1996}).

\bibitem[{\citenamefont{Gull et~al.}(2008{\natexlab{b}})\citenamefont{Gull,
  Werner, Parcollet, and Troyer}}]{Gull08_ctaux}
\bibinfo{author}{\bibfnamefont{E.}~\bibnamefont{Gull}},
  \bibinfo{author}{\bibfnamefont{P.}~\bibnamefont{Werner}},
  \bibinfo{author}{\bibfnamefont{O.}~\bibnamefont{Parcollet}},
  \bibnamefont{and} \bibinfo{author}{\bibfnamefont{M.}~\bibnamefont{Troyer}},
  \bibinfo{journal}{EPL (Europhysics Letters)} \textbf{\bibinfo{volume}{82}},
  \bibinfo{pages}{57003} (\bibinfo{year}{2008}{\natexlab{b}}).

\bibitem[{\citenamefont{Alvarez et~al.}(2008)\citenamefont{Alvarez, Summers,
  Maxwell, Eisenbach, Meredith, Larkin, Levesque, Maier, Kent, D'Azevedo
  et~al.}}]{Alvarez08}
\bibinfo{author}{\bibfnamefont{G.}~\bibnamefont{Alvarez}},
  \bibinfo{author}{\bibfnamefont{M.~S.} \bibnamefont{Summers}},
  \bibinfo{author}{\bibfnamefont{D.~E.} \bibnamefont{Maxwell}},
  \bibinfo{author}{\bibfnamefont{M.}~\bibnamefont{Eisenbach}},
  \bibinfo{author}{\bibfnamefont{J.~S.} \bibnamefont{Meredith}},
  \bibinfo{author}{\bibfnamefont{J.~M.} \bibnamefont{Larkin}},
  \bibinfo{author}{\bibfnamefont{J.}~\bibnamefont{Levesque}},
  \bibinfo{author}{\bibfnamefont{T.~A.} \bibnamefont{Maier}},
  \bibinfo{author}{\bibfnamefont{P.~R.~C.} \bibnamefont{Kent}},
  \bibinfo{author}{\bibfnamefont{E.~F.} \bibnamefont{D'Azevedo}},
  \bibnamefont{et~al.}, in \emph{\bibinfo{booktitle}{SC '08: Proceedings of the
  2008 ACM/IEEE Conference on Supercomputing}} (\bibinfo{publisher}{IEEE
  Press}, \bibinfo{address}{Piscataway, NJ, USA}, \bibinfo{year}{2008}), pp.
  \bibinfo{pages}{1--10}, ISBN \bibinfo{isbn}{978-1-4244-2835-9}.

\bibitem[{\citenamefont{Troyer and Wiese}(2005)}]{Troyer05}
\bibinfo{author}{\bibfnamefont{M.}~\bibnamefont{Troyer}} \bibnamefont{and}
  \bibinfo{author}{\bibfnamefont{U.-J.} \bibnamefont{Wiese}},
  \bibinfo{journal}{Phys. Rev. Lett.} \textbf{\bibinfo{volume}{94}},
  \bibinfo{pages}{170201} (\bibinfo{year}{2005}).

\bibitem[{\citenamefont{Andersen et~al.}(1995)\citenamefont{Andersen,
  Liechtenstein, Jepsen, and Paulsen}}]{Andersen95}
\bibinfo{author}{\bibfnamefont{O.~K.} \bibnamefont{Andersen}},
  \bibinfo{author}{\bibfnamefont{A.~I.} \bibnamefont{Liechtenstein}},
  \bibinfo{author}{\bibfnamefont{O.}~\bibnamefont{Jepsen}}, \bibnamefont{and}
  \bibinfo{author}{\bibfnamefont{F.}~\bibnamefont{Paulsen}},
  \bibinfo{journal}{J. Phys. Chem. Solids} \textbf{\bibinfo{volume}{56}},
  \bibinfo{pages}{1573 } (\bibinfo{year}{1995}).

\bibitem[{\citenamefont{Koga et~al.}(2005)\citenamefont{Koga, Kawakami, Rice,
  and Sigrist}}]{Koga05}
\bibinfo{author}{\bibfnamefont{A.}~\bibnamefont{Koga}},
  \bibinfo{author}{\bibfnamefont{N.}~\bibnamefont{Kawakami}},
  \bibinfo{author}{\bibfnamefont{T.~M.} \bibnamefont{Rice}}, \bibnamefont{and}
  \bibinfo{author}{\bibfnamefont{M.}~\bibnamefont{Sigrist}},
  \bibinfo{journal}{Phys. Rev. B} \textbf{\bibinfo{volume}{72}},
  \bibinfo{eid}{045128} (\bibinfo{year}{2005}).

\bibitem[{\citenamefont{Jakobi et~al.}(2009)\citenamefont{Jakobi, Bl\"{u}mer,
  and van Dongen}}]{Jakobi09}
\bibinfo{author}{\bibfnamefont{E.}~\bibnamefont{Jakobi}},
  \bibinfo{author}{\bibfnamefont{N.}~\bibnamefont{Bl\"{u}mer}},
  \bibnamefont{and} \bibinfo{author}{\bibfnamefont{P.}~\bibnamefont{van
  Dongen}}, \bibinfo{journal}{Phys. Rev. B} \textbf{\bibinfo{volume}{80}},
  \bibinfo{pages}{115109} (\bibinfo{year}{2009}).

\bibitem[{\citenamefont{Macridin
  et~al.}(2006{\natexlab{b}})\citenamefont{Macridin, Jarrell, and
  Maier}}]{Macridin06b}
\bibinfo{author}{\bibfnamefont{A.}~\bibnamefont{Macridin}},
  \bibinfo{author}{\bibfnamefont{M.}~\bibnamefont{Jarrell}}, \bibnamefont{and}
  \bibinfo{author}{\bibfnamefont{T.}~\bibnamefont{Maier}},
  \bibinfo{journal}{Phys. Rev. B} \textbf{\bibinfo{volume}{74}},
  \bibinfo{pages}{085104} (\bibinfo{year}{2006}{\natexlab{b}}).

\bibitem[{\citenamefont{Wang et~al.}(2009)\citenamefont{Wang, Gull, de' Medici,
  Capone, and Millis}}]{Wang09a}
\bibinfo{author}{\bibfnamefont{X.}~\bibnamefont{Wang}},
  \bibinfo{author}{\bibfnamefont{E.}~\bibnamefont{Gull}},
  \bibinfo{author}{\bibfnamefont{L.}~\bibnamefont{de' Medici}},
  \bibinfo{author}{\bibfnamefont{M.}~\bibnamefont{Capone}}, \bibnamefont{and}
  \bibinfo{author}{\bibfnamefont{A.~J.} \bibnamefont{Millis}},
  \bibinfo{journal}{Phys. Rev. B} \textbf{\bibinfo{volume}{80}},
  \bibinfo{eid}{045101} (\bibinfo{year}{2009}).

\bibitem[{\citenamefont{Pomeranchuk}(1959)}]{Pomeranchuk59}
\bibinfo{author}{\bibfnamefont{I.~I.} \bibnamefont{Pomeranchuk}},
  \bibinfo{journal}{JETP} \textbf{\bibinfo{volume}{8}}, \bibinfo{pages}{361}
  (\bibinfo{year}{1959}).

\bibitem[{\citenamefont{Oganesyan et~al.}(2001)\citenamefont{Oganesyan,
  Kivelson, and Fradkin}}]{Oganesyan01}
\bibinfo{author}{\bibfnamefont{V.}~\bibnamefont{Oganesyan}},
  \bibinfo{author}{\bibfnamefont{S.~A.} \bibnamefont{Kivelson}},
  \bibnamefont{and} \bibinfo{author}{\bibfnamefont{E.}~\bibnamefont{Fradkin}},
  \bibinfo{journal}{Phys. Rev. B} \textbf{\bibinfo{volume}{64}},
  \bibinfo{pages}{195109} (\bibinfo{year}{2001}).

\bibitem[{\citenamefont{Kivelson et~al.}(2003)\citenamefont{Kivelson, Bindloss,
  Fradkin, Oganesyan, Tranquada, Kapitulnik, and Howald}}]{Kivelson03}
\bibinfo{author}{\bibfnamefont{S.~A.} \bibnamefont{Kivelson}},
  \bibinfo{author}{\bibfnamefont{I.~P.} \bibnamefont{Bindloss}},
  \bibinfo{author}{\bibfnamefont{E.}~\bibnamefont{Fradkin}},
  \bibinfo{author}{\bibfnamefont{V.}~\bibnamefont{Oganesyan}},
  \bibinfo{author}{\bibfnamefont{J.~M.} \bibnamefont{Tranquada}},
  \bibinfo{author}{\bibfnamefont{A.}~\bibnamefont{Kapitulnik}},
  \bibnamefont{and} \bibinfo{author}{\bibfnamefont{C.}~\bibnamefont{Howald}},
  \bibinfo{journal}{Rev. Mod. Phys.} \textbf{\bibinfo{volume}{75}},
  \bibinfo{pages}{1201} (\bibinfo{year}{2003}).

\bibitem[{\citenamefont{Kim and Kee}(2004)}]{Kee04}
\bibinfo{author}{\bibfnamefont{Y.~B.} \bibnamefont{Kim}} \bibnamefont{and}
  \bibinfo{author}{\bibfnamefont{H.-Y.} \bibnamefont{Kee}},
  \bibinfo{journal}{J. Phys.: Condens. Matter} \textbf{\bibinfo{volume}{16}},
  \bibinfo{pages}{3139} (\bibinfo{year}{2004}).

\bibitem[{\citenamefont{Kee and Kim}(2005)}]{Kee05}
\bibinfo{author}{\bibfnamefont{H.-Y.} \bibnamefont{Kee}} \bibnamefont{and}
  \bibinfo{author}{\bibfnamefont{Y.~B.} \bibnamefont{Kim}},
  \bibinfo{journal}{Phys. Rev. B} \textbf{\bibinfo{volume}{71}},
  \bibinfo{pages}{184402} (\bibinfo{year}{2005}).

\bibitem[{\citenamefont{Borzi et~al.}(2007)\citenamefont{Borzi, Grigera,
  Farrell, Perry, Lister, Lee, Tennant, Maeno, and Mackenzie}}]{Borzi07}
\bibinfo{author}{\bibfnamefont{R.~A.} \bibnamefont{Borzi}},
  \bibinfo{author}{\bibfnamefont{S.~A.} \bibnamefont{Grigera}},
  \bibinfo{author}{\bibfnamefont{J.}~\bibnamefont{Farrell}},
  \bibinfo{author}{\bibfnamefont{R.~S.} \bibnamefont{Perry}},
  \bibinfo{author}{\bibfnamefont{S.~J.~S.} \bibnamefont{Lister}},
  \bibinfo{author}{\bibfnamefont{S.~L.} \bibnamefont{Lee}},
  \bibinfo{author}{\bibfnamefont{D.~A.} \bibnamefont{Tennant}},
  \bibinfo{author}{\bibfnamefont{Y.}~\bibnamefont{Maeno}}, \bibnamefont{and}
  \bibinfo{author}{\bibfnamefont{A.~P.} \bibnamefont{Mackenzie}},
  \bibinfo{journal}{Science} \textbf{\bibinfo{volume}{315}},
  \bibinfo{pages}{214} (\bibinfo{year}{2007}).

\bibitem[{\citenamefont{Fradkin et~al.}(2007)\citenamefont{Fradkin, Kivelson,
  and Oganesyan}}]{Fradkin07}
\bibinfo{author}{\bibfnamefont{E.}~\bibnamefont{Fradkin}},
  \bibinfo{author}{\bibfnamefont{S.~A.} \bibnamefont{Kivelson}},
  \bibnamefont{and}
  \bibinfo{author}{\bibfnamefont{V.}~\bibnamefont{Oganesyan}},
  \bibinfo{journal}{Science} \textbf{\bibinfo{volume}{315}},
  \bibinfo{pages}{196} (\bibinfo{year}{2007}).

\bibitem[{\citenamefont{Wu et~al.}(2007)\citenamefont{Wu, Sun, Fradkin, and
  Zhang}}]{Wu07}
\bibinfo{author}{\bibfnamefont{C.}~\bibnamefont{Wu}},
  \bibinfo{author}{\bibfnamefont{K.}~\bibnamefont{Sun}},
  \bibinfo{author}{\bibfnamefont{E.}~\bibnamefont{Fradkin}}, \bibnamefont{and}
  \bibinfo{author}{\bibfnamefont{S.-C.} \bibnamefont{Zhang}},
  \bibinfo{journal}{Phys. Rev. B} \textbf{\bibinfo{volume}{75}},
  \bibinfo{pages}{115103} (\bibinfo{year}{2007}).

\bibitem[{\citenamefont{Dell'Anna and Metzner}(2007)}]{DellAnna07}
\bibinfo{author}{\bibfnamefont{L.}~\bibnamefont{Dell'Anna}} \bibnamefont{and}
  \bibinfo{author}{\bibfnamefont{W.}~\bibnamefont{Metzner}},
  \bibinfo{journal}{Phys. Rev. Lett.} \textbf{\bibinfo{volume}{98}},
  \bibinfo{pages}{136402} (\bibinfo{year}{2007}).

\bibitem[{\citenamefont{Halboth and Metzner}(2000)}]{Halboth00}
\bibinfo{author}{\bibfnamefont{C.~J.} \bibnamefont{Halboth}} \bibnamefont{and}
  \bibinfo{author}{\bibfnamefont{W.}~\bibnamefont{Metzner}},
  \bibinfo{journal}{Phys. Rev. Lett.} \textbf{\bibinfo{volume}{85}},
  \bibinfo{pages}{5162} (\bibinfo{year}{2000}).

\bibitem[{\citenamefont{Hankevych et~al.}(2002)\citenamefont{Hankevych, Grote,
  and Wegner}}]{Hankevych02}
\bibinfo{author}{\bibfnamefont{V.}~\bibnamefont{Hankevych}},
  \bibinfo{author}{\bibfnamefont{I.}~\bibnamefont{Grote}}, \bibnamefont{and}
  \bibinfo{author}{\bibfnamefont{F.}~\bibnamefont{Wegner}},
  \bibinfo{journal}{Phys. Rev. B} \textbf{\bibinfo{volume}{66}},
  \bibinfo{pages}{094516} (\bibinfo{year}{2002}).

\bibitem[{\citenamefont{Yamase and Kohno}(2000{\natexlab{a}})}]{Yamase00}
\bibinfo{author}{\bibfnamefont{H.}~\bibnamefont{Yamase}} \bibnamefont{and}
  \bibinfo{author}{\bibfnamefont{H.}~\bibnamefont{Kohno}}, \bibinfo{journal}{J.
  Phys. Soc. Jpn.} \textbf{\bibinfo{volume}{69}}, \bibinfo{pages}{332}
  (\bibinfo{year}{2000}{\natexlab{a}}).

\bibitem[{\citenamefont{Yamase and Kohno}(2000{\natexlab{b}})}]{Yamase00B}
\bibinfo{author}{\bibfnamefont{H.}~\bibnamefont{Yamase}} \bibnamefont{and}
  \bibinfo{author}{\bibfnamefont{H.}~\bibnamefont{Kohno}}, \bibinfo{journal}{J.
  Phys. Soc. Jpn.} \textbf{\bibinfo{volume}{69}}, \bibinfo{pages}{2151}
  (\bibinfo{year}{2000}{\natexlab{b}}).

\bibitem[{\citenamefont{Edegger et~al.}(2006)\citenamefont{Edegger, Muthukumar,
  and Gros}}]{Edegger06}
\bibinfo{author}{\bibfnamefont{B.}~\bibnamefont{Edegger}},
  \bibinfo{author}{\bibfnamefont{V.~N.} \bibnamefont{Muthukumar}},
  \bibnamefont{and} \bibinfo{author}{\bibfnamefont{C.}~\bibnamefont{Gros}},
  \bibinfo{journal}{Phys. Rev. B} \textbf{\bibinfo{volume}{74}},
  \bibinfo{pages}{165109} (\bibinfo{year}{2006}).

\bibitem[{\citenamefont{Werner and Millis}(2007)}]{Werner072level}
\bibinfo{author}{\bibfnamefont{P.}~\bibnamefont{Werner}} \bibnamefont{and}
  \bibinfo{author}{\bibfnamefont{A.~J.} \bibnamefont{Millis}},
  \bibinfo{journal}{Phys. Rev. Lett.} \textbf{\bibinfo{volume}{99}},
  \bibinfo{pages}{126405} (\bibinfo{year}{2007}).

\bibitem[{\citenamefont{Werner et~al.}(2009{\natexlab{b}})\citenamefont{Werner,
  Gull, and Millis}}]{Werner093level}
\bibinfo{author}{\bibfnamefont{P.}~\bibnamefont{Werner}},
  \bibinfo{author}{\bibfnamefont{E.}~\bibnamefont{Gull}}, \bibnamefont{and}
  \bibinfo{author}{\bibfnamefont{A.~J.} \bibnamefont{Millis}},
  \bibinfo{journal}{Phys. Rev. B} \textbf{\bibinfo{volume}{79}},
  \bibinfo{pages}{115119} (\bibinfo{year}{2009}{\natexlab{b}}).

\bibitem[{\citenamefont{Hufner et~al.}(2008)\citenamefont{Hufner, Hossain,
  Damascelli, and Sawatzky}}]{Hufner08}
\bibinfo{author}{\bibfnamefont{S.}~\bibnamefont{Hufner}},
  \bibinfo{author}{\bibfnamefont{M.~A.} \bibnamefont{Hossain}},
  \bibinfo{author}{\bibfnamefont{A.}~\bibnamefont{Damascelli}},
  \bibnamefont{and} \bibinfo{author}{\bibfnamefont{G.~A.}
  \bibnamefont{Sawatzky}}, \bibinfo{journal}{Rep. Prog. Phys.}
  \textbf{\bibinfo{volume}{71}}, \bibinfo{pages}{062501}
  (\bibinfo{year}{2008}).

\bibitem[{\citenamefont{Khatami et~al.}(2009)\citenamefont{Khatami, Mikelsons,
  Galanakis, Macridin, Moreno, Scalettar, and Jarrell}}]{Khatami09}
\bibinfo{author}{\bibfnamefont{E.}~\bibnamefont{Khatami}},
  \bibinfo{author}{\bibfnamefont{K.}~\bibnamefont{Mikelsons}},
  \bibinfo{author}{\bibfnamefont{D.}~\bibnamefont{Galanakis}},
  \bibinfo{author}{\bibfnamefont{A.}~\bibnamefont{Macridin}},
  \bibinfo{author}{\bibfnamefont{J.}~\bibnamefont{Moreno}},
  \bibinfo{author}{\bibfnamefont{R.~T.} \bibnamefont{Scalettar}},
  \bibnamefont{and} \bibinfo{author}{\bibfnamefont{M.}~\bibnamefont{Jarrell}},
  \bibinfo{journal}{unpublished, preprint arXiv:0909.0759}
  (\bibinfo{year}{2009}).

\bibitem[{\citenamefont{Albuquerque et~al.}(2007)\citenamefont{Albuquerque,
  Alet, Corboz et~al.}}]{ALPS}
\bibinfo{author}{\bibfnamefont{A.}~\bibnamefont{Albuquerque}},
  \bibinfo{author}{\bibfnamefont{F.}~\bibnamefont{Alet}},
  \bibinfo{author}{\bibfnamefont{P.}~\bibnamefont{Corboz}},
  \bibnamefont{et~al.}, \bibinfo{journal}{Journal of Magnetism and Magnetic
  Materials} \textbf{\bibinfo{volume}{310}}, \bibinfo{pages}{1187}
  (\bibinfo{year}{2007}).

\end{thebibliography}

\end{document}